\documentclass[final]{article}
\usepackage{latexsym, graphicx, epsfig, amsmath, amsfonts,amssymb,subfigure,algorithm,algorithmicx,booktabs,bm,color,bigints}
\usepackage{amsthm}

\def\w{{\omega}}

\newcommand{\pd}[2]{\frac{\partial#1}{\partial#2}}

\usepackage{authblk}

\begin{document}

\title{Model reduction for a power grid model}

\author[]{Jing Li and Panos Stinis}

\affil[]{Advanced Computing, Mathematics and Data Division, Pacific Northwest National Laboratory, Richland WA 99354}

\renewcommand\Affilfont{\itshape\small}

\date {}

\maketitle

\begin{abstract}
We apply model reduction techniques to the DeMarco power grid model. The DeMarco model, when augmented by an appropriate line failure mechanism, can be used to study cascade failures. Here we examine the DeMarco model without the line failure mechanism and we investigate how to construct reduced order models for subsets of the state variables. We show that due to the oscillating nature of the solutions and the absence of timescale separation between resolved and unresolved variables, the construction of accurate reduced models becomes highly non-trivial since one has to account for long memory effects. In addition, we show that a reduced model which includes even a short memory is drastically better than a memoryless model.  
\end{abstract}

\section{Introduction}

The importance of stability of the power grid cannot be overstated (see e.g. \cite{dobson2007,simonsen2008,tamrakar2018,sloothaak2018,motter2013,witthaut2012} for a collection of recent articles on various aspects of power grid stability). The study of stability of the power grid usually involves large scale computations. With this in mind, in the current work we examine the feasibility of accurate reduced order models for systems of equations describing power grid dynamics. In particular, we focus on the DeMarco power grid model \cite{zheng2010}. The DeMarco model, when augmented by an appropriate line failure mechanism, can be used to study cascade failures. Here we examine the DeMarco model without the line failure mechanism and we investigate how to construct reduced order models for subsets of the state variables. 

Given the oscillatory nature of the solutions of power grid models, we expect that there is no clear timescale separation between most of the state variables. Indeed, this turns out to be the case. This situation renders the construction and subsequent simulation of a reduced model to be non-trivial tasks. In particular, there are long memory effects \cite{givon2004,CS06}. This means that constructing an accurate reduced model for a subset of the variables, called resolved variables, requires to account for the history of those variables. In fact, we show that if we want the reduced model to retain its accuracy, the length of the history that we need to account for turns out to be equal to the interval of time we wish to evolve the reduced model for. This is one of the most challenging cases of model reduction because we have to find a way to represent accurately such long memory effects. 

We employ the Mori-Zwanzig (MZ) formalism (see e.g. \cite{CHK00}) which allows in principle the construction of reduced order models with any memory length. This facilitates the investigation of the effect of truncating the memory length as well as the cost needed to obtain an accurate representation of the memory. Our results for the DeMarco model are informative. First, as mentioned above, accurate prediction by a reduced model requires the length of the memory to be equal to the time interval we wish to obtain a prediction for. Second, even if we retain a long memory, the simulation of the integro-differential equations of the reduced model (the integral part is due to the memory) requires an adequately small timestep to remain stable. Third, if we truncate the memory length, the reduced model loses accuracy fast for time intervals longer than the truncated memory length, but it does not lose its stability as long as we use a small enough timestep. Fourth, even the inclusion of a short memory is better than a memoryless reduced model.  Finally, at least for the initial conditions examined here, we observe that while the DeMarco model is nonlinear, the memory is accurately represented using a {\it linear} function of the resolved variables. This is interesting because it can lead to simplification of the reduced model which is not obvious at first sight.

Motivated by the encouraging results we obtained for the DeMarco model, we used a simple example of a single particle coupled linearly to a heat bath of linear oscillators put forth by Zwanzig (see e.g. Section 1.6 in \cite{zwanzig2001}), where the exact reduced model can be constructed analytically, to put our results in context.  For this simple example we found the same qualitative features in the behavior of the reduced model as in the reduced model for the considerably more complicated DeMarco model. In this sense, the results we obtained for the reduced order model of the DeMarco model are optimal even though we had to perform certain approximations to facilitate the computations. 

The paper is organized as follows. In Section \ref{mz_formalism} we present the MZ formalism and explain our choice of a projection operator. In Section \ref{demarco_power_grid} we present the $n$-bus DeMarco model as well as the 3-bus version that we used for our numerical experiments. Section \ref{numerical} contains the construction of the reduced order model,  results of its simulation as well as numerical results for the exact reduced model for the simple example of a single particle coupled linearly to a heat bath of linear oscillators. Finally, Section \ref{discussion} contains a short discussion of the results and suggestions for future work.

\section{The Mori-Zwanzig formalism}\label{mz_formalism}

We begin in Section \ref{mz_formalism_general} with a brief presentation of the Mori-Zwanzig (MZ) formalism for constructing reduced models of systems of differential equations (see \cite{CHK00,CHK3,CS06} for more details). The MZ formalism belongs in the class of projection methods. As such, the choice of projection operator can affect significantly the final form of the reduced models. Section \ref{mz_formalism_projection} contains a discussion of various projection operators used in the literature as well as the choice we have made in the current work.

\subsection{The Mori-Zwanzig equation}\label{mz_formalism_general}

Suppose we are given the system
\begin{equation}\label{odes}
\frac{du(t)}{dt} = R (t,u(t)),
\end{equation}
where $u = ( \{u_k\}), \; k \in H \cup G$
with initial condition $u(0)=u_0.$ The unknown variables (modes) are divided into two groups, one group is indexed in H and the order indexed in G. Our goal is to construct a reduced model for the modes in the set $H.$ The system of ordinary differential equations
we are given can be transformed into a system of  linear
partial differential equations
\begin{equation}
\label{pde}
\pd{\varphi_k}{t}=L \varphi_k, \qquad \varphi_k (u_0,0)=u_{0k}, \, k \in H \cup G
\end{equation}
where $L=\sum_{k \in H \cup G } R_i(u_0) \frac{\partial}{\partial u_{0i}}.$ The solution of \eqref{pde} is
given by $u_k (u_0,t)=\varphi_k(u_0,t)$. Using semigroup notation we can rewrite (\ref{pde}) as
$$\pd{}{t} e^{tL} u_{0k}=L e^{tL} u_{0k}$$
Suppose that the vector of initial conditions can be divided as $u_0=(\hat{u}_0,\tilde{u}_0),$ where
$\hat{u}_0$ is the vector of the resolved variables (those in $H$) and $\tilde{u}_0$ is the vector of the unresolved variables (those in $G$).  Let $P$ be an orthogonal projection on the space of functions of $\hat{u}_0$ and $Q=I-P$ (see also Section \ref{mz_formalism_projection} for a more detailed discussion of projection operators).

Equation \eqref{pde}
can be rewritten as
\begin{equation}
\label{mz}
\frac{\partial}{\partial{t}} e^{tL}u_{0k}=
e^{tL}PLu_{0k}+e^{tQL}QLu_{0k}+
\int_0^t e^{(t-s)L}PLe^{sQL}QLu_{0k}ds, \, k \in H,
\end{equation}
where we have used Dyson's formula
\begin{equation}
\label{dyson1}
e^{tL}=e^{tQL}+\int_0^t e^{(t-s)L}PLe^{sQL}ds.
\end{equation}
Equation (\ref{mz}) is the Mori-Zwanzig identity.
Note that
this relation is exact and is an alternative way
of writing the original PDE. It is the starting
point of our approximations. Of course, we
have one such equation for each of the resolved
variables $u_k, k \in H$. The first term in (\ref{mz}) is
usually called Markovian since it depends only on the values of the variables
at the current instant, the second is called ``noise" and the third ``memory" (see \cite{CS06} for a discussion of the significance of each term).

If we write
$$e^{tQL}QLu_{0k}=w_k,$$
$w_k(u_0,t)$ satisfies the equation
\begin{equation}
\label{ortho}
\begin{cases}
&\frac{\partial}{\partial{t}}w_k(u_0,t)=QLw_k(u_0,t) \\
& w_k(u_0,0) = QLu_{0k}=R_k(u_0)-(PR_k)(\hat{u_0}).
\end{cases}
\end{equation}
If we project (\ref{ortho}) we get
$$P\frac{\partial}{\partial{t}}w_k(u_0,t)=
PQLw_k(u_0,t)=0,$$
since $PQ=0$. Also for the initial condition
$$Pw_k(u_0,0)=PQLu_{0k}=0$$
by the same argument. Thus, the solution
of (\ref{ortho}) is at all times orthogonal
to the range of $P.$ We call
(\ref{ortho}) the orthogonal dynamics equation (see more details in \cite{CHK00}). Since the solutions of
the orthogonal dynamics equation remain orthogonal to the range of $P$,
we can project the Mori-Zwanzig equation (\ref{mz}) and find
\begin{equation}
\label{mzp}
\frac{\partial}{\partial{t}} Pe^{tL}u_{0k}=
Pe^{tL}PLu_{0k}+
P\int_0^t e^{(t-s)L}PLe^{sQL}QLu_{0k} ds.
\end{equation}
We will not present here more details about how to start from Eq. \eqref{mzp} and construct reduced models of different orders for a general system of ODEs. Such constructions have been documented thoroughly elsewhere (see e.g. \cite{CHK3}). However, we will provide such details for the specific example of the power grid equations in Section \ref{numerical_3_bus_reduced}.

\subsection{The choice of the projection operator $P$}\label{mz_formalism_projection}

Before we proceed we would like to comment on choices of the projection operator $P$ that have appeared in the literature as well as the choice that we have opted for in the current work. As we have seen in Eq. \eqref{mz}, what the MZ formalism offers is a way to decompose the RHS of the equations for the resolved variables. This decomposition involves three terms. The choice of the projection operator affects how the information content of the RHS of the equation is distributed among these three terms. The popular choices for the projection operator (see e.g. \cite{CHK3}) include the conditional expectation and the finite-rank projection with respect to an {\it invariant} measure for the system. The projection operator based on the conditional expectation also comes with the added property of being optimal in an $L_2$ sense. One uses these projection operators when the objective is to produce trajectories for the resolved variables starting from initial conditions sampled from the invariant measure or when one wants to study the relaxation to the invariant measure when starting from a non-typical initial condition.

For the power grid model that we have studied, we do not have access to an invariant measure. The projection operator $P$ we have chosen is defined as  $(Pf)(\hat{u}_0) = P(f(u_0))=P(f(\hat{u}_0,\tilde{u}_0))=f(\hat{u}_0,\tilde{u}^0).$ What this definition means is that our projection operator, when applied to a function of the initial conditions, assigns the value $\tilde{u}^0$ to the unresolved variables, where $\tilde{u}^0$ is a {\it chosen} vector of values for the unresolved variables. One way to think about this that makes contact with other choices for the projection operator that have appeared in the literature, is that we have used a projection operator which is defined with respect to a measure on the initial conditions that has a delta measure for the unresolved variables centered at $\tilde{u}^0.$ Thus, it does {\it not} allow fluctuations for the initial condition of the unresolved variables. This particular choice of projection operator with $\tilde{u}^0=\tilde{0}$ has been used successfully by the current authors before to tackle a variety of problems from detection and tracking of singularities to locating bifurcations and uncertainty quantification (see e.g. \cite{bib2,li2016,li2019,bib3,bib7}). 

Our choice of operator comes with a few distinctive features. First, unlike the more popular choices of the projection operator presented above, our chosen operator commutes with nonlinear functions. This property turns out to be convenient for the analytical calculations of some, but not all, of the expressions that appear in the MZ formalism. Second, for the case $\tilde{u}^0=\tilde{0},$ our choice of projection operator is the Galerkin projection that sets the unresolved variables to zero for all time. However, note that our reduced models incorporate the memory terms too which are there to account for the interaction of the resolved and unresolved variables. Third, since our choice of projection operator does {\it not} allow fluctuations in the initial conditions of the unresolved variables, then when we apply the projection operator to the MZ equation to cancel the noise term, we get an equation which is valid {\it pathwise}. In other words, the projected, {\it noiseless} MZ equation \eqref{mzp} is valid for the prediction of {\it each} trajectory that starts with whatever initial conditions we have chosen for the resolved variables and $\tilde{u}^0$ for the unresolved variables. Fourth, as we have stated earlier in this paragraph, the specific choice of the projection operator $P$ allows the analytical calculations of some, but not all, of the expressions in the MZ formalism.
We will return to this point in Section \ref{3_bus_reduced_memory}.


\section{The DeMarco model for the power grid}\label{demarco_power_grid}

\subsection{The $n$-bus system}\label{n_bus}

The state of the $n$-bus DeMarco model for the power grid can be described in terms of a vector $u=(\omega_g, \delta_g, \delta_l, V_l)^T$ where $\omega_g$ contains the angular velocities of the generators, $\delta_g, \delta_l$ contain the bus voltage phase angles with respect to an arbitrary synchronous reference frame of the generators and loads respectively and $V_l$ contains the voltage magnitudes of the loads.

We define the energy function $\Phi$ by
\begin{equation} \label{demarco_energy_function}
\Phi(u)=\frac{1}{2}\omega_g^T M_g \omega_g + (P_g + P_g^a) \cdot \delta_g + (P_l + P_l^a) \cdot \delta_l + (Q_g + Q_l^a) \cdot \ln V_l
\end{equation}
where $(\ln V_l)_i=\ln V_{li}.$ Also, $P_g, P_l$ is the active power injected into the generator and load buses respectively, $Q_l$ is the reactive power injected into the load buses and $P_g^a, P_l^a, Q_l^a $ is the power absorbed by the generator and load buses respectively.

The gradient of $\Phi$ is given by 
\begin{equation} \label{demarco_energy_function_gradient}
 \nabla_u \Phi (u)= \left [\begin{array}{r}
 M_g \omega_g \\
P_g + P_g^a \\
P_l + P_l^a\\
  V_l^{-1}(Q_g + Q_l^a)\\
   \end{array} \right ] 
\end{equation}     
where $(V_l^{-1})_i=V_{li}^{-1}.$ We also define the full-rank, negative semi-definite matrix $A$ given by
\begin{equation}\label{demarco_matrix_A}
A = \left [\begin{array}{rrrr}
 -M_g^{-1}D_gM_g^{-1} &-M_g^{-1} & 0 & 0\\
M_g^{-1} &  0 & 0 & 0 \\
0&  0 & -D_l^{-1} & 0\\
  0&0&0&-\frac{1}{\epsilon} * I\\
   \end{array} \right ] 
\end{equation}        
where $M_g,D_g,D_l, \epsilon$ are system parameters and $I$ is the identity matrix. With the definitions \eqref{demarco_energy_function}-\eqref{demarco_matrix_A}, the dynamic equations for the DeMarco model are given by
\begin{equation}\label{demarco_dynamics}
\frac{du}{dt}=A \nabla_u \Phi(u). 
\end{equation} 
Given the negative semi-definiteness of the matrix $A,$ we see that 
\begin{equation}\label{demarco_stability}
\frac{\partial \Phi}{\partial t}=\nabla_u \Phi (u) \cdot \frac{du}{dt}= \nabla_u \Phi^T(u) A \nabla_u \Phi (u) \leq 0.
\end{equation} 
We note that DeMarco model was developed to study the problem of cascading failure and its energy function contains an additional term to account for the possible line failure between nodes \cite{zheng2010}. In the current work we are not interested in constructing reduced models for a system where line failures may occur, so we omit this extra term from the energy function. Reduced models for system with possible line failures will be described in a future publication.


\subsection{The 3-bus system}\label{3_bus}
\begin{figure}[htbp]
\centerline{
\includegraphics[width = 11cm]{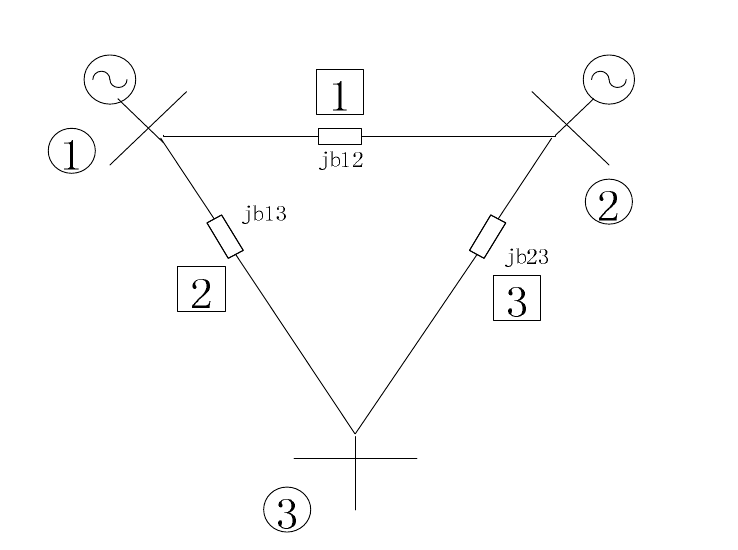}
}
\caption{A 3-bus system.}
\label{fig:3bus}
\end{figure}

In this section we focus on the $n=3$ case, i.e. a 3-bus system (see \cite{zheng2010} and Fig. \ref{fig:3bus}). In particular, Bus 1 is the reference generator.  The voltage magnitude $V_1$ and the voltage phase angle $\delta_1$ are fixed while the injected powers $P_1$ and $Q_1$ vary according to the system. Bus 2 is a regular generator. The voltage magnitude $V_2$ and the active power injected $P_2$ are fixed while the reactive power $Q_2$ and voltage phase angle $\delta_2$ vary. Bus 3 is a loading bus. The injected powers $P_3$ and $Q_3$ are fixed while the voltage magnitude $V3$ and voltage phase angle $\delta_3$ vary. With this arrangement, there are five variables in the system: $\omega_1$, $\omega_2$, $\alpha_2 = \delta_2 -\delta_1$, $\alpha_3 = \delta_3-\delta_1$, $V_3.$ Thus, $u= (\omega_1,\omega_2,\alpha_2,\alpha_3,V_3)^{T}$.

The energy function $\Phi_{3bus}(u)$ is 

\begin{gather}
\Phi_{3bus} (u) = \frac{1}{2}M_1 \omega_1^2 + \frac{1}{2}M_2 \omega_2^2+\frac{1}{2}(b_1+b_2)V_1^2+\frac{1}{2}(b_1+b_3)V_2^2 \notag \\
+\frac{1}{2}(b_2+b_3)V_3^2-b_1 V_1 V_2 \cos(\alpha_2) -b_2 V_1 V_3\cos(\alpha_3)  \notag \\
-b_3V_2 V_3 \cos(\alpha_3-\alpha_2)+P_2\alpha_2+P_3\alpha_3+Q_3\ln(V_3) \label{3_bus_energy_function}
\end{gather}

The gradient $\nabla_u \Phi_{3bus}(u)=[\frac{\partial \Phi_{3bus}}{\partial \omega_1}, \frac{\partial \Phi_{3bus}}{\partial \omega_2}, \frac{\partial \Phi_{3bus}}{\partial \alpha_2}, \frac{\partial \Phi_{3bus}}{\partial \alpha_3}, \frac{\partial \Phi_{3bus}}{\partial V_3} ]^T$ is given by
\begin{eqnarray}
\frac{\partial \Phi_{3bus}}{\partial \omega_1} &=& M_1 \omega_1, \label{3_bus_energy_function_gradient1}\\
\frac{\partial \Phi_{3bus}}{\partial \omega_2} &=& M_2 \omega_2, \label{3_bus_energy_function_gradient2}\\
\frac{\partial \Phi_{3bus}}{\partial \alpha_2} &=& b_1V_1V_2\sin(\alpha_2) + b_3V_2V_3\sin(\alpha_2-\alpha_3)+P_2, \label{3_bus_energy_function_gradient3} \\
\frac{\partial \Phi_{3bus}}{\partial \alpha_3} &=& b_2V_1V_3\sin(\alpha_3) + b_3V_2V_3\sin(\alpha_3-\alpha_2)+P_3, \label{3_bus_energy_function_gradient4}\\
\frac{\partial \Phi_{3bus}}{\partial V_3} &=& (b_2+b_3)V_3 - b_2V_1\cos(\alpha_3)-b_3 V_2\cos(\alpha_3-\alpha_2)+\frac{Q_3}{V_3}. \label{3_bus_energy_function_gradient5}
\end{eqnarray}
The matrix $A$ is given by 
\begin{equation}
A = \left [\begin{array}{rrrrr}
 -M_1^{-1}D_1M_1^{-1} & 0 &M_1^{-1} &M_1^{-1} & 0\\
0 &  -M_2^{-1}D_2M_2^{-1} &-M_2^{-1} & 0 & 0 \\
 -M_1^{-1} &  M_2^{-1} & 0 & 0 & 0\\
  -M_1^{-1} &0 & 0& -D_3^{-1} &0\\
  0&0&0&0&-\epsilon^{-1}\\
   \end{array} \right ]. \label{3_bus_matrix_A}
 \end{equation}  
The values of the parameters $M_1,D_1,M_2,D_2,D_3$ and $\epsilon$ used for the numerical experiments are provided in Table \ref{tab:para}. 

\begin{table}
\begin{center}
\begin{tabular}{ccccccc}
\hline
$M_1$ & $M_2$ & $b_1$ & $b_2$ & $b_3$ & $D_1$ & $D_2$ \\
$0.052$ & $0.0531$ & $10$ & $10$ & $10$ & $0.05$ & $0.05$ \\
\hline
 $D_3$ & $P_2$ & $P_3$ & $Q_3$ & $\epsilon$ & $V_1$ & $V_2$ \\
 $0.005$ & $-2.0$ & $3.0$ & $0.1$ & $5$ & $0.9$ & $0.9$ \\
\hline
\end{tabular}
\end{center}
\caption{Parameters for the 3-bus system}
\label{tab:para}
\end{table}

With the definitions \eqref{3_bus_energy_function}-\eqref{3_bus_matrix_A}, the dynamics for the 3-bus system are given by
\begin{equation}\label{3_bus_system}
\frac{du}{dt} = A\nabla_{u}\Phi_{3bus}(u).
\end{equation}

The power flow equations are given by
\begin{eqnarray}
P_i &=& \sum_j B_{i,j}V_i V_j \sin(\delta_i-\delta_j) \;  \textrm{for all buses } \;(i= 1,2,3) \label{3_bus_power_flow1} \\
Q_i  &=& -\sum_j B_{i,j} V_iV_j\cos(\delta_i-\delta_j) \;  \textrm{for loading buses } \; (i = 3) \label{3_bus_power_flow2}
\end{eqnarray}
where $B_{i,j}$ are parameters. Note that with appropriately chosen $B_{i,j}$, the fixed-point equations $\nabla_{u}\Phi_{3bus}(u) = 0$ coincide with the power flow equations.

\section{Numerical results}\label{numerical}

We want to examine the behavior of MZ reduced order models for the case of the DeMarco model of the power grid. We focus on the case of a 3-bus system. In Section \ref{numerical_3_bus_full} we present results for the full order 3-bus system which helps us show the {\it absence} of time scale separation between the different variables of the system. In Section \ref{numerical_3_bus_reduced} we present numerical results for the MZ reduced order model of the 3-bus system when we resolve only 3 of the 5 variables of the system. 

\subsection{Full order system}\label{numerical_3_bus_full}

The 3-bus system was simulated with the forward Euler scheme with initial condition given by 
\begin{equation}\label{3_bus_full_initial} 
\omega_1(0) = 0, \quad \omega_2(0) = 0, \quad  \alpha_2(0) = -0.16, \quad \alpha_3(0)=-0.3 \quad V_3(0) = 0.8.   
\end{equation}
The values for the parameters of the 3-bus system are listed in Table \ref{tab:para}. In this study, we choose the variables associated with the generators as the resolved variables, and the variables associated with the loads as the unresolved variables. As a result, the unresolved variables are $\alpha_3$ and $V_3$. 

\begin{figure}[htbp]
   \centering
   \subfigure[]{%
   \includegraphics[width = 3.5cm]{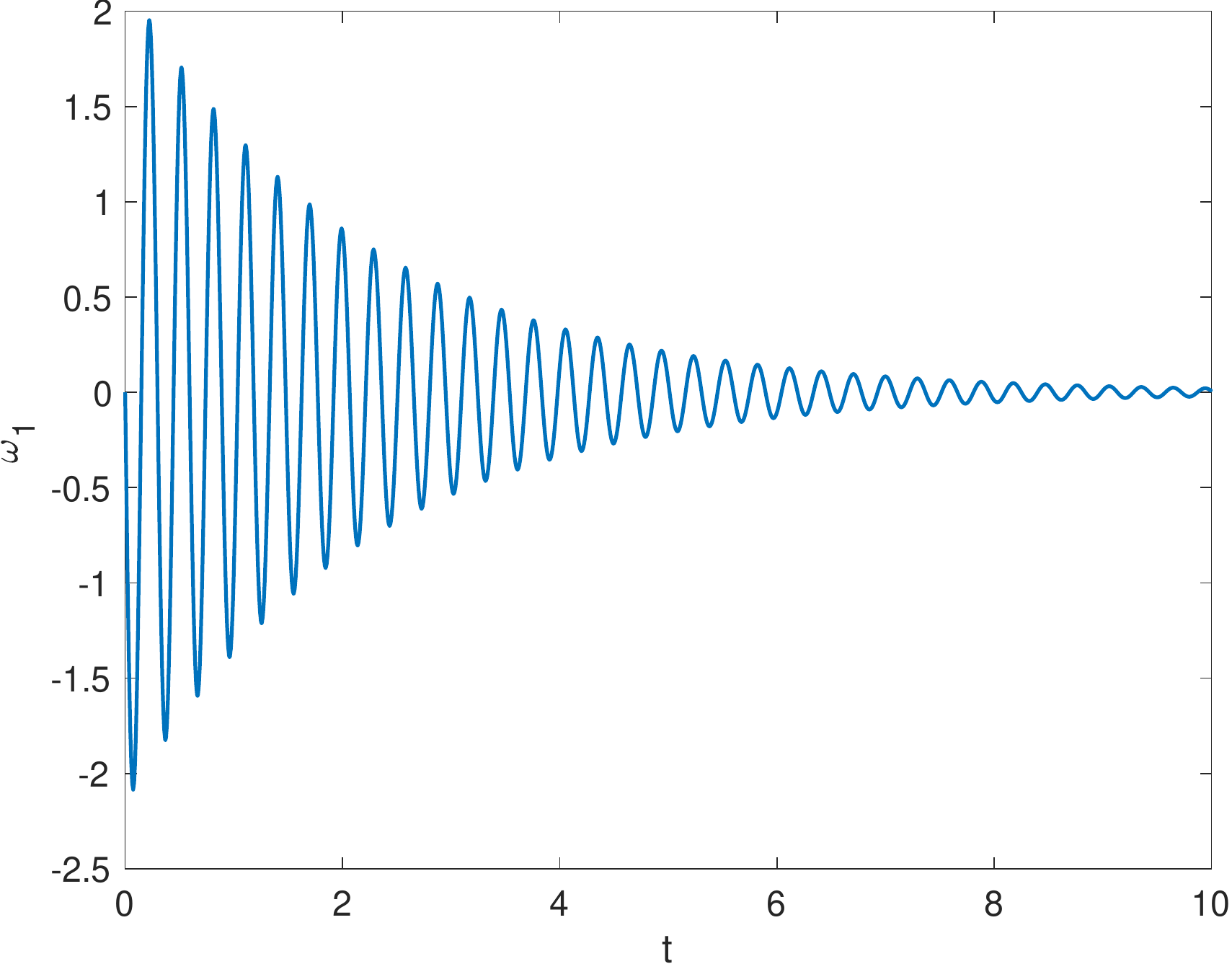}
   \label{fig:w1}}
      \quad
   \subfigure[]{%
   \includegraphics[width = 3.5cm]{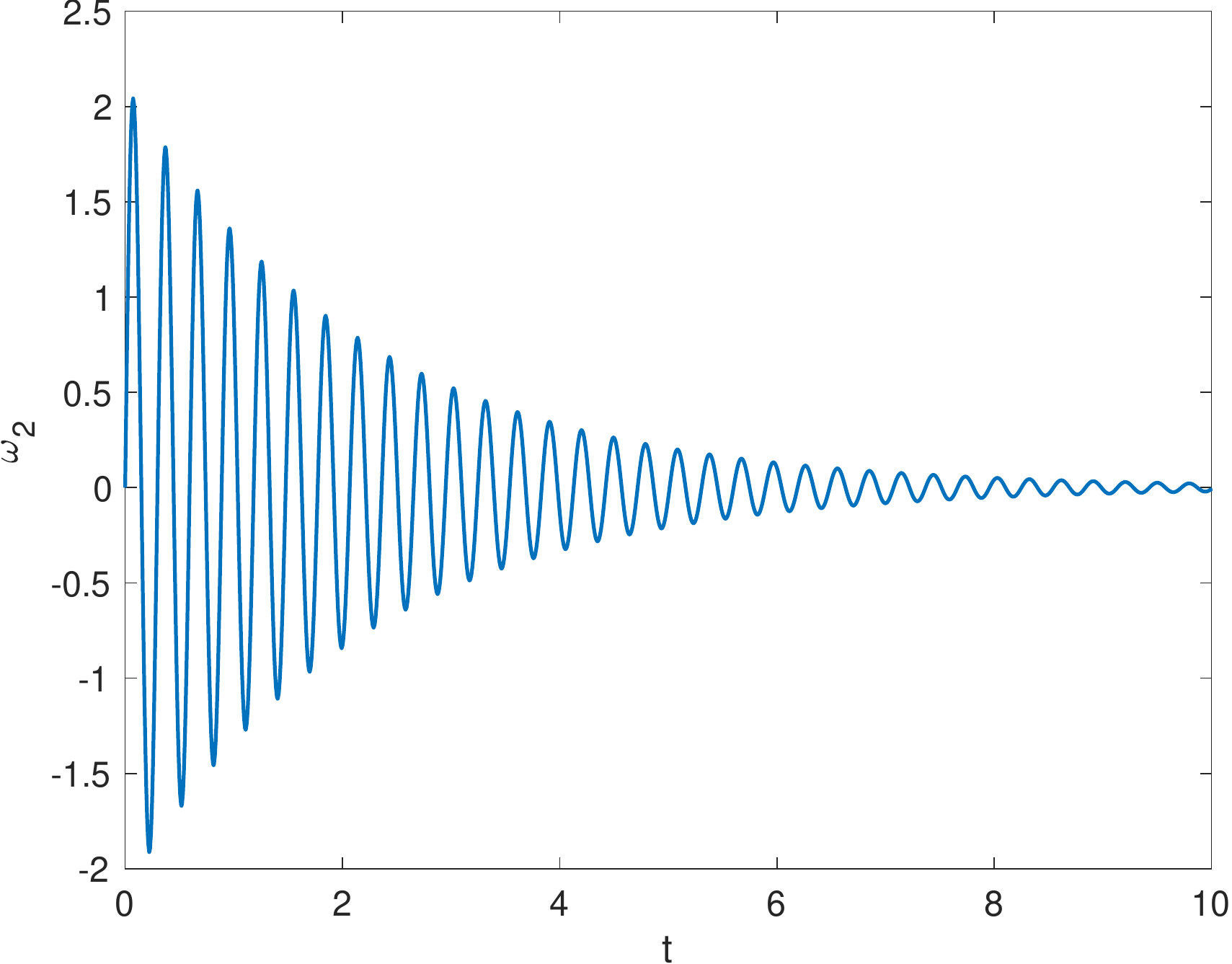}
   \label{fig:w2}}
      \quad
   \subfigure[]{%
   \includegraphics[width = 3.5cm]{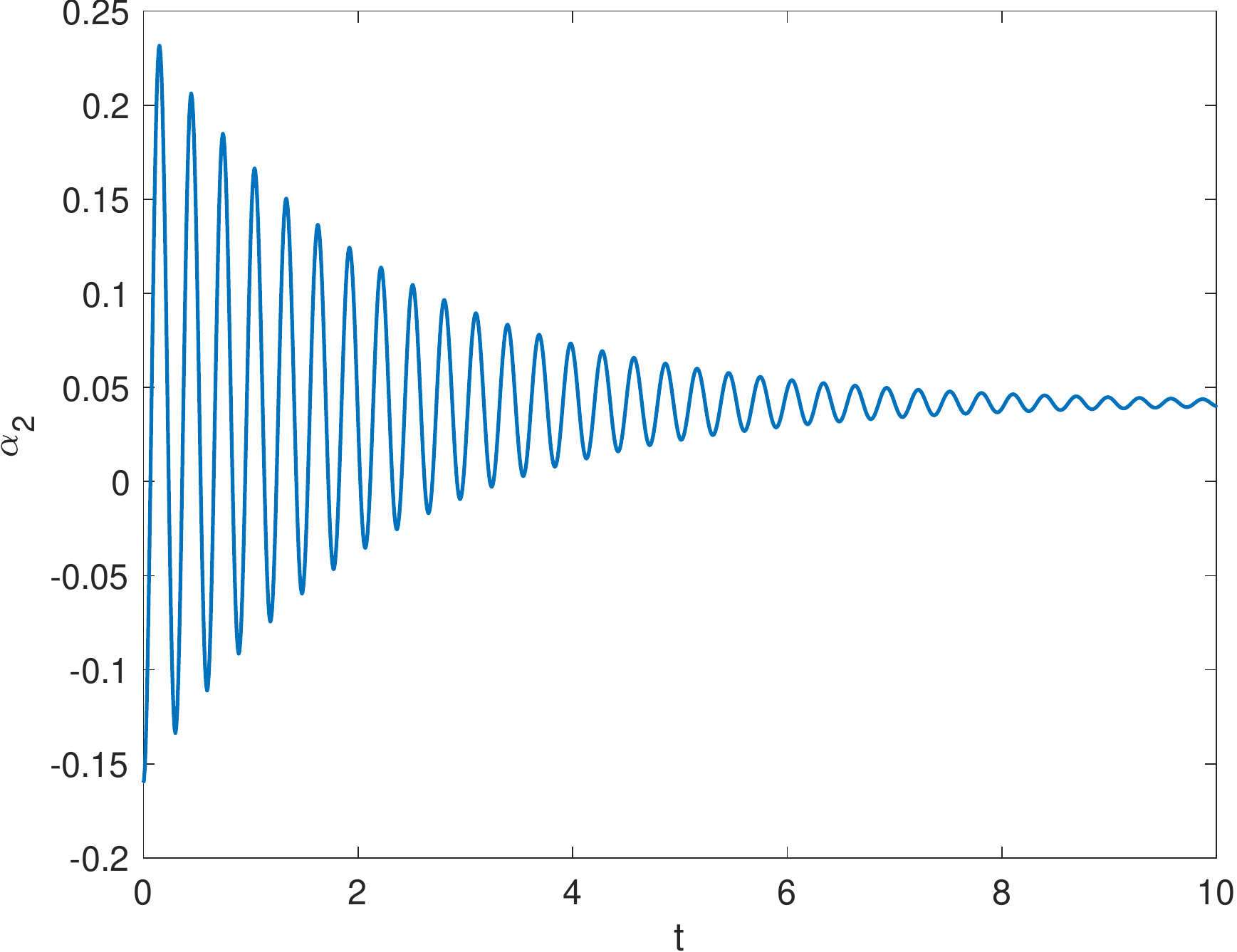}
   \label{fig:a2}}
\caption{Full system - Evolution of {\it resolved} variables. a) $\omega_1$ for generator 1, b) $\omega_2$ for generator 2 and c) $\alpha_2$ for generator 2 .}
\label{fig:generator_2}
\end{figure}

\begin{figure}[htbp]
   \centering
   \subfigure[]{%
   \includegraphics[width = 5cm]{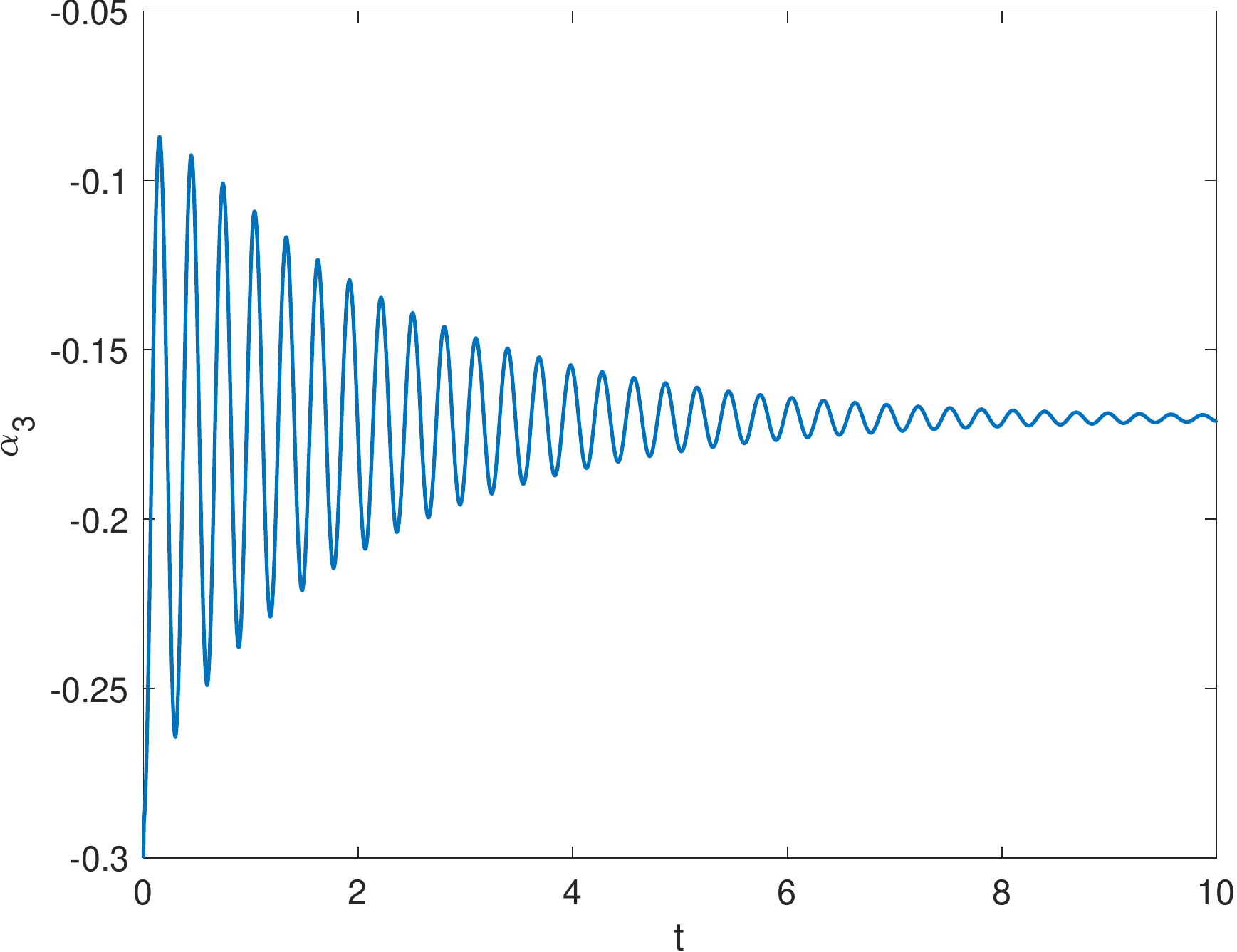}
   \label{fig:a3}}
      \quad
   \subfigure[]{%
   \includegraphics[width = 5cm]{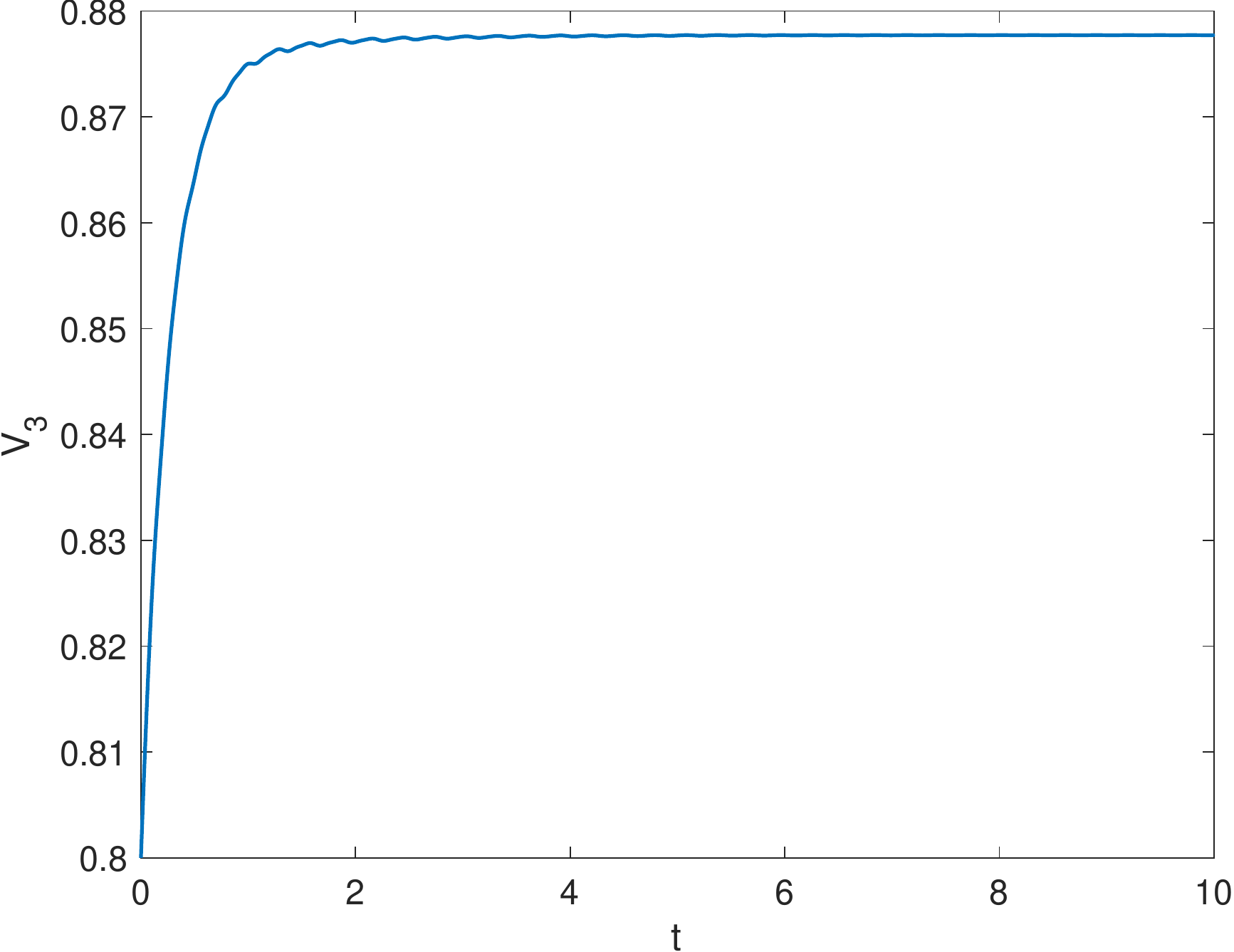}
   \label{fig:V3}}
\caption{Full system - Evolution of {\it unresolved} variables. a) $\alpha_3$ for load bus 3, b) $V_3$ for load bus 3.}
\label{fig:unresolved}
\end{figure}

Figures \ref{fig:generator_2} and \ref{fig:unresolved} show the evolution of quantities associated with the second generator (resolved) and the load (unresolved) respectively. The purpose of showing these results is to establish that there is a {\it absence} of timescale separation in the evolution of the resolved and unresolved variables. This is very important for the construction of the reduced model since it leads to long memory effects. In particular, by examining Fig. \ref{fig:a2} and Fig. \ref{fig:a3} for the evolution of the resolved variable $\alpha_2$ and the unresolved variable $\alpha_3$ respectively, we see that they oscillate on very similar timescales. As a result, we expect that the memory term will be oscillating for long time instead of decaying quickly as in problems with timescale separation. In Section \ref{numerical_3_bus_reduced} we will establish i) that indeed the memory is long and ii) that naive treatment of such long memory effects leads to rapid loss of accuracy of the reduced model. 

\subsection{Reduced system with 3 resolved variables}\label{numerical_3_bus_reduced}

\subsubsection{The MZ formalism}\label{3_bus_reduced_mz}
To derive the MZ reduced model for the 3-bus system, we first rewrite it in the notation of the MZ formalism given in Section \ref{mz_formalism}. As stated in Section \ref{numerical_3_bus_full}, we choose the variables associated with the generators as the resolved variables, and those associated with the loads as the unresolved variables. As a result, the unresolved variables are $\alpha_3$ and $V_3.$ As explained in Section \ref{mz_formalism_projection}, for any function $f(\omega_{10},\omega_{20},\alpha_{20},\alpha_{30},V_{30})$ of the initial conditions of all the variables, we define the projection operator $P$ as $Pf(\omega_{10},\omega_{20},\alpha_{20},\alpha_{30},V_{30})= f(\omega_{10},\omega_{20},\alpha_{20},\alpha_3^0,V_3^0).$ 
This means that we assign to the unresolved variables $\alpha_3$ and $V_3$ the values $\alpha_3^0$ and $V_3^0.$ There is no variance allowed for the values of the unresolved variables. We note that in our numerical examples we also tested the case of allowing the unresolved to have a small variance around $\alpha_3^0$ and $V_3^0$ , but found no significant change in the performance of the reduced MZ models. We decided to use the projection operator which does not allow any variance since it facilitates the derivation of expressions needed for the MZ reduced models. 

Even if we change the initial values for the {\it resolved} variables, the projection operator will assign to the {\it unresolved} variables the same values $\alpha_3^0$ and $V_3^0.$ This is a Galerkin-type projection but instead of assigning to the unresolved variables the value 0, we assign $\alpha_3^0$ and $V_3^0.$ However, note the important difference that in a Galerkin model, the unresolved variables would be assigned the same value for {\it all} times. As we have explained in Section \ref{mz_formalism_projection}, we allow the unresolved variables to evolve and their interaction with the resolved variables is accounted for through the memory term.      

Let 
\begin{eqnarray}
R_1 &=& a_{1,1}M_1\w_1+a_{1,3}[b_1V_1V_2\sin(\alpha_2)+b_3V_2V_3\sin(\alpha_2-\alpha_3)+P_2] \notag\\
&+&a_{1,4}[b_2V_1V_3\sin(\alpha_3) + b_3V_2V_3\sin(\alpha_3-\alpha_2)+P_3], \label{3_bus_rhs1} \\
R_2  &=& a_{2,2}M_2\w_2+a_{2,3}[b_1V_1V_2\sin(\alpha_2)+b_3V_2V_3\sin(\alpha_2-\alpha_3)+P_2], \label{3_bus_rhs2}\\
R_3  &=& a_{3,1}M_1\w_1+a_{3,2}M_2\w_2, \label{3_bus_rhs3}\\
R_4  &=& a_{4,1}M_1\w_1+a_{4,4}[b_2V_1V_3\sin(\alpha_3) + b_3V_2V_3\sin(\alpha_3-\alpha_2)+P_3], \label{3_bus_rhs4} \\
R_5  &=& a_{5,5}[(b_2+b_3)V_3 - b_2V_1\cos(\alpha_3)-b_3 V_2\cos(\alpha_3-\alpha_2)+\frac{Q_3}{V_3}], \label{3_bus_rhs5}
\end{eqnarray}
where $a_{i,j}$ is the $(i,j)$-th element of the matrix $A$ (see Eq. \eqref{3_bus_matrix_A}). Using \eqref{3_bus_rhs1}-\eqref{3_bus_rhs5}, the system of equations \eqref{3_bus_system} for the 3-bus system can be rewritten as
\begin{equation}\label{3_bus_system_rewrite}
\frac{du}{dt} = R(u).
\end{equation}

\subsubsection{The Markovian term}\label{3_bus_reduced_markovian}

The system \eqref{3_bus_system_rewrite} is the starting point for the MZ reduced models. Recall that in the MZ formalism all expressions are computed at $t=0$ i.e. using the initial conditions (see Eq. \eqref{mz}). The Markovian terms for the MZ reduced model equations for the 3 resolved variables $\hat{u}=(\omega_{1},\omega_{2},\alpha_{2})^T$ are given by 

\begin{gather}
PL\omega_{10}=PR_1(u_0) = \mathcal{R}_1(\hat{u}_0)=a_{1,1}M_1\w_{10}+a_{1,3}[b_1V_1V_2\sin(\alpha_{20}) \notag  \\+b_3V_2V_3^o\sin(\alpha_{20}-\alpha_3^o)+P_2] \notag  \\
+a_{1,4}[b_2V_1V_3^o\sin(\alpha_3^o) + b_3V_2V_3^o\sin(\alpha_3^o-\alpha_{20})+P_3], \label{3_bus_rhs1_projected} \\
PL\omega_{20}=PR_2(u_0) = \mathcal{R}_2(\hat{u}_0)= a_{2,2}M_2\w_{20}+a_{2,3}[b_1V_1V_2\sin(\alpha_{20}) \notag  \\+b_3V_2V_3^o\sin(\alpha_{20}-\alpha_3^o) + P_2], \label{3_bus_rhs2_projected} \\
PL\alpha_{20}=PR_3(u_0) = \mathcal{R}_3(\hat{u}_0)= a_{3,1}M_1\w_{10}+a_{3,2}M_2\w_{20}. \label{3_bus_rhs3_projected}
\end{gather}

\subsubsection{The memory term}\label{3_bus_reduced_memory}

We continue with the presentation of some details for the expressions in the memory term. We define the inner product
\begin{equation}
(f,g) = \int fg d\rho.
\end{equation}
where $\rho$ is the joint probability measure with respect to the initial conditions for all the variables in the full system. This can be any measure we choose and for the 3-bus system we have chosen it to be a Gaussian for a very specific purpose which we now explain. The projection operator $P$ we have used sets the initial conditions for the unresolved variables equal to some pre-chosen values. In order to compute the memory kernel we utilize a finite-rank projection operator $\mathbb{P}$ which approximates $P.$ Since the original operator $P$ allows zero variance for the unresolved variables, we have chosen $\mathbb{P}$ to be defined through a measure $\rho$ which assigns a Gaussian distribution of small variance for the initial condition the resolved variables while fixing the unresolved variables to the chosen initial conditions. In our numerical experiments for the 3-bus system we took the variance to be $10^{-4}$ for the resolved variables. The mean of this Gaussian distribution for each resolved variable is equal to the initial value of this variable. In other words, the choice of the measure $\rho$ does not have to respect any invariance for the system dynamics. If one has access to such an invariant measure, it can be used for the definition of a finite-rank projection operator (see e.g. \cite{CHK3}).

For a function $\varphi_j(u_0,t)$ of the initial conditions and time, the finite-rank projection reads (see e.g. \cite{CHK3})
\begin{equation}\label{def:frank_proj}
(\mathbb{P}\varphi_j)(\hat{u}_0,t) = \sum_{\nu\in I}(\varphi_j(u_0,t),h^{\nu}(\hat{u}_0))h^{\nu}(\hat{u}_0),
\end{equation}
where $h^{\nu}(\hat{u}_0)$ are tensor product Hermite polynomials up to some order $p$, $\nu$ is the multi-index $\nu = (\nu_1,\nu_2,\nu_3)$ with $|\nu| = \sum_{i=1}^3 \nu_i$ and $I$ is the index set up to order $p$, i.e., $I = \{ \mu \big| |\mu|\leq p \}$. For the 3-bus system, the highest order $p$ that we consider for the basis functions is 5 for a total of $\frac{(3+5)!}{3!5!}=56$ basis functions. 

For each $j\leq 3$, the component $F_j(u_0,t)$ denotes the solution of the orthogonal dynamics
\begin{equation}\label{Orth_j}
\begin{split}
&\frac{\partial}{\partial t}F_j(u_0,t) ={Q}LF_j(u_0,t) = {L}F_j(u_0,t)-{P}LF_j(u_0,t),\\
& F_j(u_0,0) = {Q}Lu_{0j} = R_j(u_0)-{P}Lu_{0j}.
\end{split}
\end{equation}
Eq. \eqref{Orth_j} is equivalent to the Dyson formula \cite{CHK00}:
\begin{equation}\label{Orth_j_Dyson}
F_j(u_0,t) = e^{tL}F_j(u_0 ,0)-\int_{0}^t e^{(t-s)L}{P}LF_j(u_0,s)ds.
\end{equation}
Eq. \eqref{Orth_j_Dyson} is a Volterra integral equation for $F_j(u_0,t).$ To proceed, we replace the projection operator ${P}$ with the finite-rank projection operator $\mathbb{P}$ and find
\begin{equation}\label{memory_kernel}
K_j(\hat{u}_0,s) = PLF_j(u_0,s) \approx \mathbb{P}LF_j(u_0,s) =  \sum_{\nu \in I}b_j^{\nu}(s)h^{\nu}(\hat{u}_0),
\end{equation}
where
\begin{equation}\label{b_nu_def}
b^{\nu}_j(s) = (LF_j(u_0,s),h^{\nu}(\hat{u}_0)).
\end{equation}
Consequently,
\[
e^{(t-s)L}\mathbb{P}LF_j(u_0,s) = \sum_{\nu\in I}b^{\nu}_j(s)h^{\nu}(e^{(t-s)L}\hat{u}_0).
\]
We substitute $e^{(t-s)L}\mathbb{P}LF_j(u_0,s)$ for $e^{(t-s)L}PLF_j(u_0,s)$ in Eq. \eqref{Orth_j_Dyson}, multiply both sides by $L$ and take the inner product with $h^{\mu}(\hat{u}_0))$; the result is (dropping the approximation sign)

\begin{equation}
\begin{split}
&(LF_j(u_0,t),h^{\mu}(\hat{u}_0)) \\
=& (Le^{tL}F_j(u_0,0),h^{\mu}(\hat{u}_0))-\int_{0}^t \sum_{\nu\in I} b^{\nu}_j(s)(Le^{(t-s)L}h^{\nu}(\hat{u}_0),h^{\mu}(\hat{u}_0))ds. \label{Volterra_a1}
\end{split}
\end{equation}
Eq. \eqref{Volterra_a1} is a Volterra integral equation for the function $b^{\nu}_j(t)$, which can be rewritten as follows:
\begin{equation}\label{Volterra_a}
b^{\mu}_j(t) = f^{\mu}_j(t)-\int_{0}^t\sum_{\nu\in I}b^{\nu}_j(s)g^{\nu\mu}(t-s)ds,
\end{equation}
where
\[
f^{\mu}_j(t) = (Le^{tL}F_j(u_0,0),h^{\mu}(\hat{u}_0)), \qquad g^{\nu\mu}(t)=(Le^{tL}h^{\nu}(\hat{u}_0),h^{\mu}(\hat{u}_0)).
\]
The functions $f^{\nu}_j(t)$, $g^{\mu\nu}(t)$ can be found by averaging over a collection of experiments or simulations, with initial conditions drawn from the initial distribution. In this example, we use a sparse grid quadrature rule for the multi-dimensional integrals \cite{xiu2005}. For the particular reduced model, we used a 3-dimensional sparse grid with a total of 681 points. 

Note that we need to clarify how the expressions $Le^{tL}F_j(u_0,0)$ and $Le^{tL}h^{\nu}(\hat{u}_0)$ that appear in the expressions for $f^{\nu}_j(t)$, $g^{\mu\nu}(t)$ can be estimated. Both of these expressions involve an application of the operator $L$ which differentiates w.r.t. to the initial conditions to expressions, $e^{tL}F_j(u_0,0)$ and $e^{tL}h^{\nu}(\hat{u}_0)$ which depend on the solution at time $t.$ Thus, they are unknown functions of the initial conditions. To proceed with the differentiation we use the formula  (see \cite{CHK00})
\begin{equation} \label{memory_22}
L G(u(u_0,t))= \sum_{r=0}^{M}R_r(u_0) \pd{}{u_{0r}}
G(u(u_0,t))=
\sum_{r=0}^{M}R_r(u(u_0,t))
(\pd{G}{u_{0r}})
(u(u_0,t)),
\end{equation}
which holds for any function $G(u(u_0,t))$ of the solution at time $t$ and where $u(u_0,t)=e^{tL}u_0.$ For example, for $Le^{tL}F_j(u_0,0)$ we find
$$Le^{tL}F_j(u_0,0)=\sum_{r=0}^{M}R_r(u(u_0,t))
(\pd{F_j}{u_{0r}})(u(u_0,t)). $$

Finally, we perform one more projection to eliminate the noise term (see Section \ref{mz_formalism}). There are two ways to compute the effect of the second projection on the memory term, in particular how to compute 
\[
\int_{0}^t Pe^{(t-s)L}K_j(\hat{u}_0,s)ds.
\]
We present both way and discuss them comparatively. The first way is to use the property of the projection operator $P$ that it commutes with a nonlinear function. Thus, from the expression $K_j(\hat{u}_0,s)=\sum_{\nu \in I}b_j^{\nu}(s)h^{\nu}(\hat{u}_0)$ (see Eq. \eqref{memory_kernel}) we get
\begin{equation}\label{second_projection_1}
\int_{0}^t Pe^{(t-s)L}K_j(\hat{u}_0,s)ds = \sum_{\nu \in I} \int_{0}^t b_j^{\nu}(s)h^{\nu}(Pe^{(t-s)L}\hat{u}_0) ds.
\end{equation}
The expression for the memory term in Eq. \eqref{second_projection_1} is in the form of a convolution sum. Thus, to evaluate it for any time $t$ we actually have to keep the history of values $h^{\nu}(Pe^{(t-s)L}\hat{u}_0)$ from time 0 to time $t$ which becomes increasingly expensive as time progresses. On the other hand, if the memory extends only to $t_{memory}$ units of time in the past, then one needs to keep only the recent $t_{memory}$ units of time of the history of $h^{\nu}(Pe^{(t-s)L}\hat{u}_0).$ The expression for the memory becomes 
\begin{equation}\label{second_projection_2}
 \int_{0}^t Pe^{(t-s)L}K_j(\hat{u}_0,s)ds = \sum_{\nu \in I} \int_{0}^{t_{memory}} b_j^{\nu}(s)h^{\nu}(Pe^{(t-s)L}\hat{u}_0)ds.
\end{equation}
The second way to compute the effect of the second projection on the memory term is to use the approximation of the projection operator $P$ by the operator $\mathbb{P}$ and the memory term becomes
\begin{equation}\label{second_projection_3}
\int_{0}^t Pe^{(t-s)L}K_j(\hat{u}_0,s)ds = \int_{0}^t \sum_{\nu\mu\in I}b^{\nu}_j(s)\gamma^{\nu\mu}(t-s)h^{\mu}(\hat{u}_0)ds,
\end{equation}
where
\begin{equation}\label{second_projection_4}
\gamma^{\nu\mu}(t) = (e^{tL}h^{\nu}(\hat{u}_0),h^{\mu}(\hat{u}_0)).
\end{equation}
After calculating $b_i^{\mu}$ and $\gamma^{\mu\nu}$ we find
\begin{equation}\label{second_projection_5}
\int_{0}^t Pe^{(t-s)L}K_j(\hat{u}_0,s)ds = \int_{0}^t B(s)\Gamma(t-s)h(\hat{u}_0)ds,
\end{equation}
where $B$ and $\Gamma$ are the matrix form of $b^{\mu}_i$ and $\gamma^{\mu\nu}$ respectively and $h(\hat{u}_0)$ the vector of basis functions. The advantage of this representation is that we can rewrite it as
\begin{equation}\label{second_projection_6}
\int_{0}^t Pe^{(t-s)L}K_j(\hat{u}_0,s)ds = \left( \int_{0}^t B(s)\Gamma(t-s) \right) h(\hat{u}_0)ds.
\end{equation}
The expression in parentheses in Eq. \eqref{second_projection_6} can be computed once offline and stored as a function of time $t.$ Then, it can be used to compute extremely efficiently the memory term for different initial conditions. The drawback of this approach is that unlike the first way,  even if the memory is finite with length $t_{memory},$ to compute $\int_{0}^{t_{memory}} B(s)\Gamma(t-s)ds $ for $t > 2t_{memory},$ we need to keep calculating $\Gamma(t).$ This calculation involves again the evolution of the full system for multiple initial conditions which is very expensive. For our example with the 3-bus system that involves an extremely long memory, we chose the second way because it makes the simulation of the reduced order model very efficient. 

With this choice for the memory term we obtain for the 3-bus system the following MZ reduced model,
\begin{equation}\label{eq:redu_b}
\frac{d}{dt}\hat{u}(t) = \hat{\mathcal{R}}(\hat{u}(t))+\int_{0}^t B(s)\Gamma(t-s)h(\hat{u}_0)ds, \quad \hat{u}(0) = \hat{u}_0.
\end{equation}
The vector $\hat{\mathcal{R}}(\hat{u}(t)) $ contains the Markovian terms \eqref{3_bus_rhs1_projected}-\eqref{3_bus_rhs3_projected} and $\hat{u}_0$ is the initial condition of the resolved variables.

\begin{figure}[htbp]
   \centering
   \subfigure[]{%
   \includegraphics[width = 5cm]{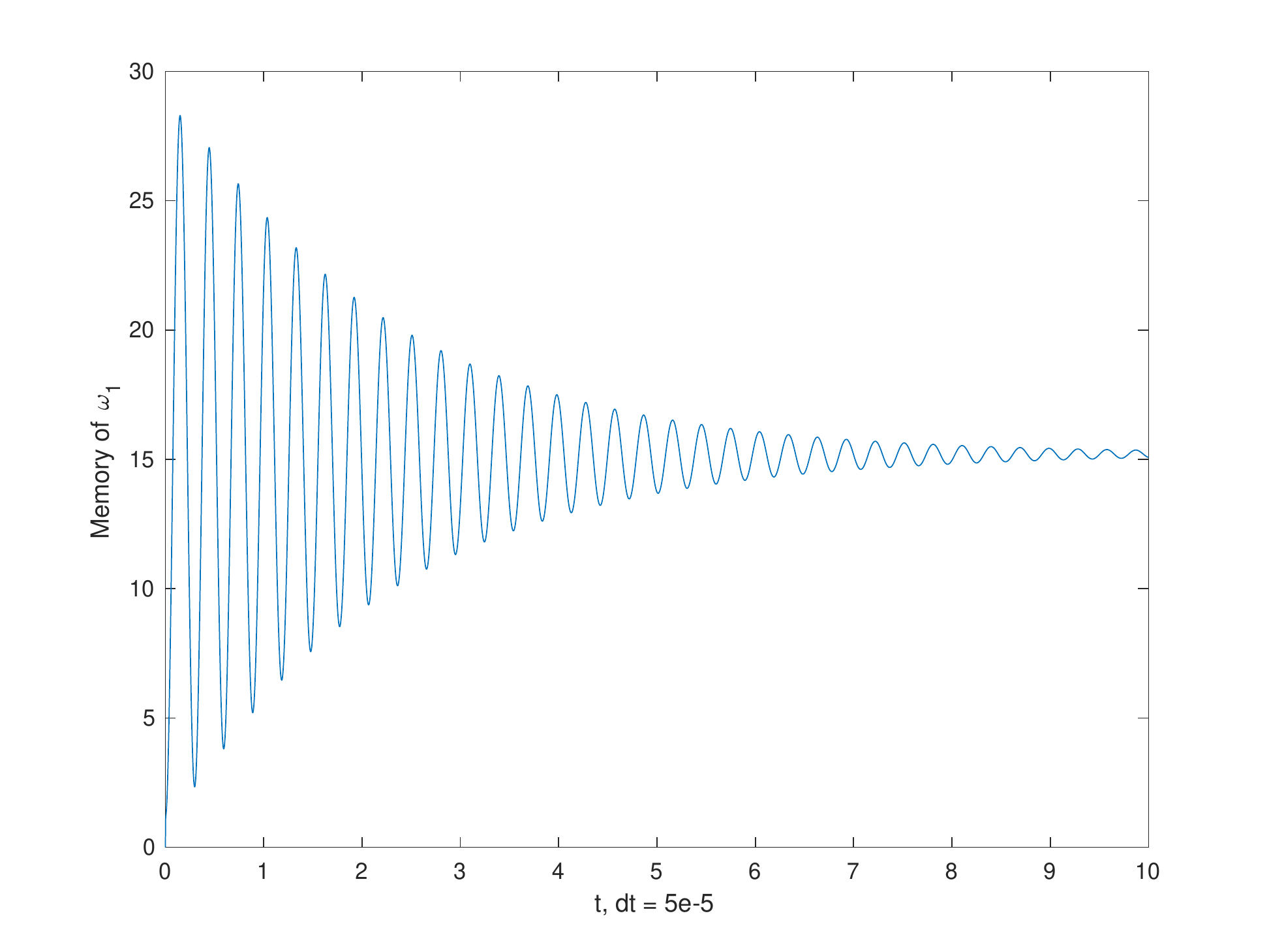}
   \label{fig:w1memory}}
      \quad
   \subfigure[]{%
   \includegraphics[width = 5cm]{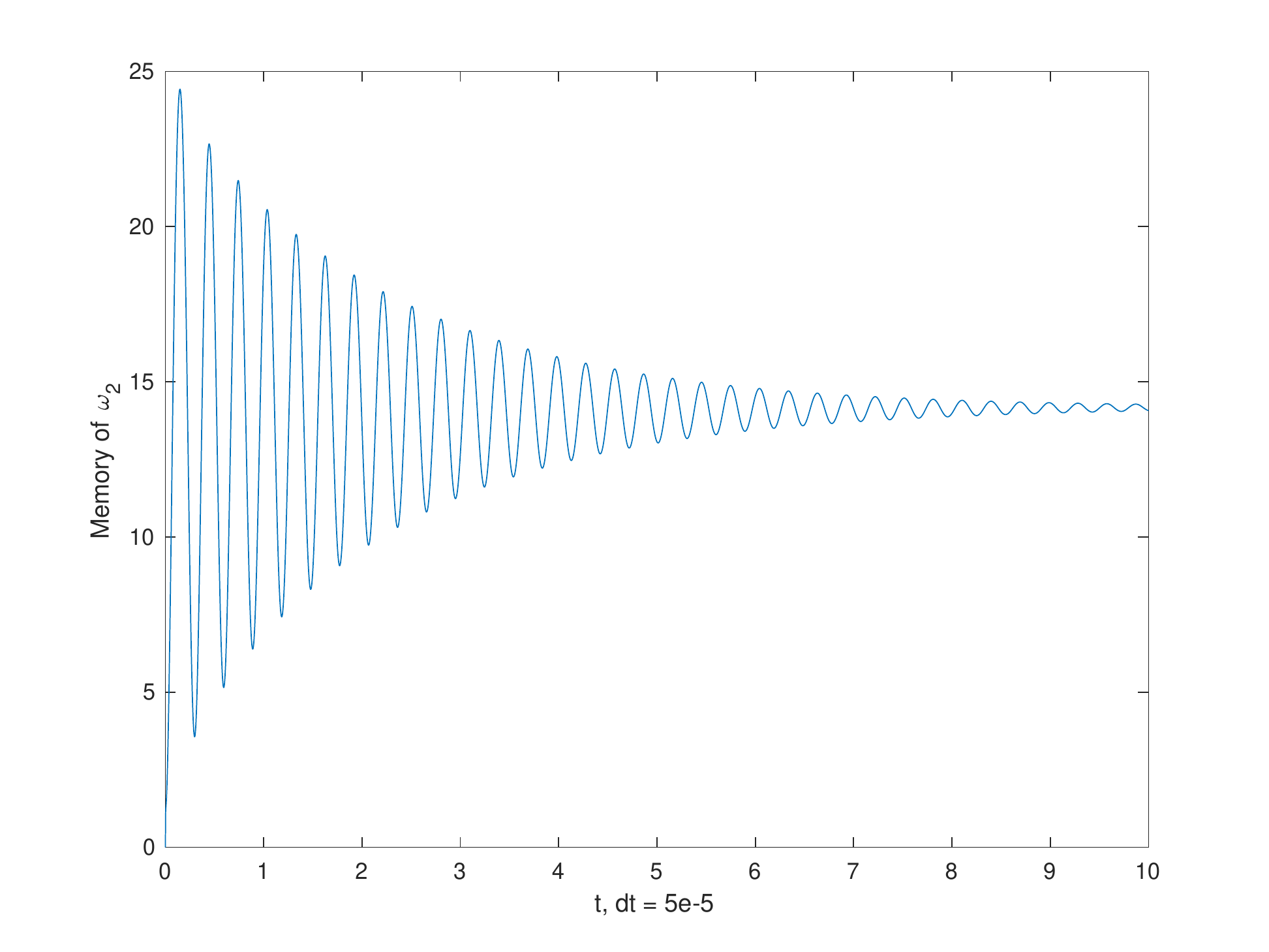}
   \label{fig:w2memory}}
\caption{Reduced system - Evolution of memory. a) For $\omega_1,$ b) For $\omega_2.$}
\label{fig:memory}
\end{figure}

Figure \ref{fig:memory} shows the evolution of the memory for $\omega_1$ and $\omega_2.$ It keeps oscillating for a long time which makes the construction of accurate reduced models more difficult.

\subsubsection{Numerical schemes for the reduced model}\label{3_bus_reduced_scheme}

After setting up the reduced model \eqref{eq:redu_b}, we need to solve it to obtain the evolution of the resolved variables. The system \eqref{eq:redu_b} contains integro-differential equations. Thus, we need to also decide how to evaluate the memory term which is given by a (convolution) integral. We tested two different numerical implementations for the evaluation of the memory term. 

The first scheme employs a forward Euler scheme both for the Markovian and the memory terms. Also, we approximate the memory integral with the trapezoidal rule. In particular, we have
\begin{equation}
\begin{split}
\hat{u}_j(t_{n+1}) = & \hat{u}_j(t_{n}) + \Delta t \mathcal{R}_j(\hat{u}(t_n)) + \frac{ \Delta t^2}{2} \sum_{i=0}^{n-1}\left(\sum_{\nu\mu\in I} b_{j}^{\nu}(t_i)\gamma^{\nu\mu}(t_n-t_i)h^{\mu}(\hat{u}_0) \right.\\
&\left. + \sum_{\nu\mu\in I} b_{j}^{\nu}(t_{i+1})\gamma^{\nu\mu}(t_n-t_{i+1})h^{\mu}(\hat{u}_0)\right)
\end{split}
\end{equation}
where $j=1,2,3$ for the 3 resolved variables $\omega_1,\omega_2,\alpha_2$ and $t_n=(n-1)\Delta t.$ The numerical results we present used and $\Delta t = 5 \times 10^{-5}$ and $\Delta t = 10^{-4}.$

The second scheme employs a forward Euler scheme for the Markovian term and an implicit Euler scheme for the memory term. Again, we approximate the memory integral with the trapezoidal rule. In particular, we have
\begin{equation}
\begin{split}
\hat{u}_j(t_{n+1}) = & \hat{u}_j(t_{n}) + \Delta t \mathcal{R}_j(\hat{u}(t_n)) + \frac{\Delta t^2}{2} \sum_{i=0}^{n}\left(\sum_{\nu\mu\in I} b_{j}^{\nu}(t_i)\gamma^{\nu\mu}(t_{n+1}-t_i)h^{\mu}(\hat{u}_0) \right.\\
&\left. + \sum_{\nu\mu\in I} b_{j}^{\nu}(t_{i+1})\gamma^{\nu\mu}(t_{n+1}-t_{i+1})h^{\mu}(\hat{u}_0)\right)
\end{split}
\end{equation}
where $j=1,2,3$ for the 3 resolved variables $\omega_1,\omega_2,\alpha_2$ and $t_{n+1}=n\Delta t.$

We did not observe large differences between the results of the two numerical schemes. We will present results only for the first scheme (forward Euler scheme for Markovian and memory terms).

%

\subsubsection{Numerical results for variable memory length}\label{3_bus_reduced_variable}
In this section we present results of the reduced model for various values of the memory length including the case {\it without} a memory term which uses only the Markovian term. The range of integration in the memory term of the system \eqref{eq:redu_b} extends from 0 to $t.$ Note that due to the convolutional nature of the integral which represents the memory term, for $s=0$ the memory kernel integrand corresponds to the current time $t$ and for $s=t$ it corresponds to the initial time 0. For system \eqref{eq:redu_b}, the memory includes {\it all} the history of the system. We call this case the {\it infinite} memory case.

For the case of {\it finite} memory length, say $t_{memory},$ we have to truncate the range of integration from 0 to $t_{memory},$ i.e. we take into account only the recent $t_{memory}$ units of the history. The system \eqref{eq:redu_b} is rewritten as
\begin{equation}\label{eq:redu_b_finite}
\frac{d}{dt}\hat{u}(t) = \hat{\mathcal{R}}(\hat{u}(t))+\int_{0}^{t_{memory}} B(s)\Gamma(t-s)h(\hat{u}_0)ds, \quad \hat{u}(0) = \hat{u}_0.
\end{equation}

\begin{figure}[htbp]
   \centering
   \subfigure[]{%
   \includegraphics[width = 3.5cm]{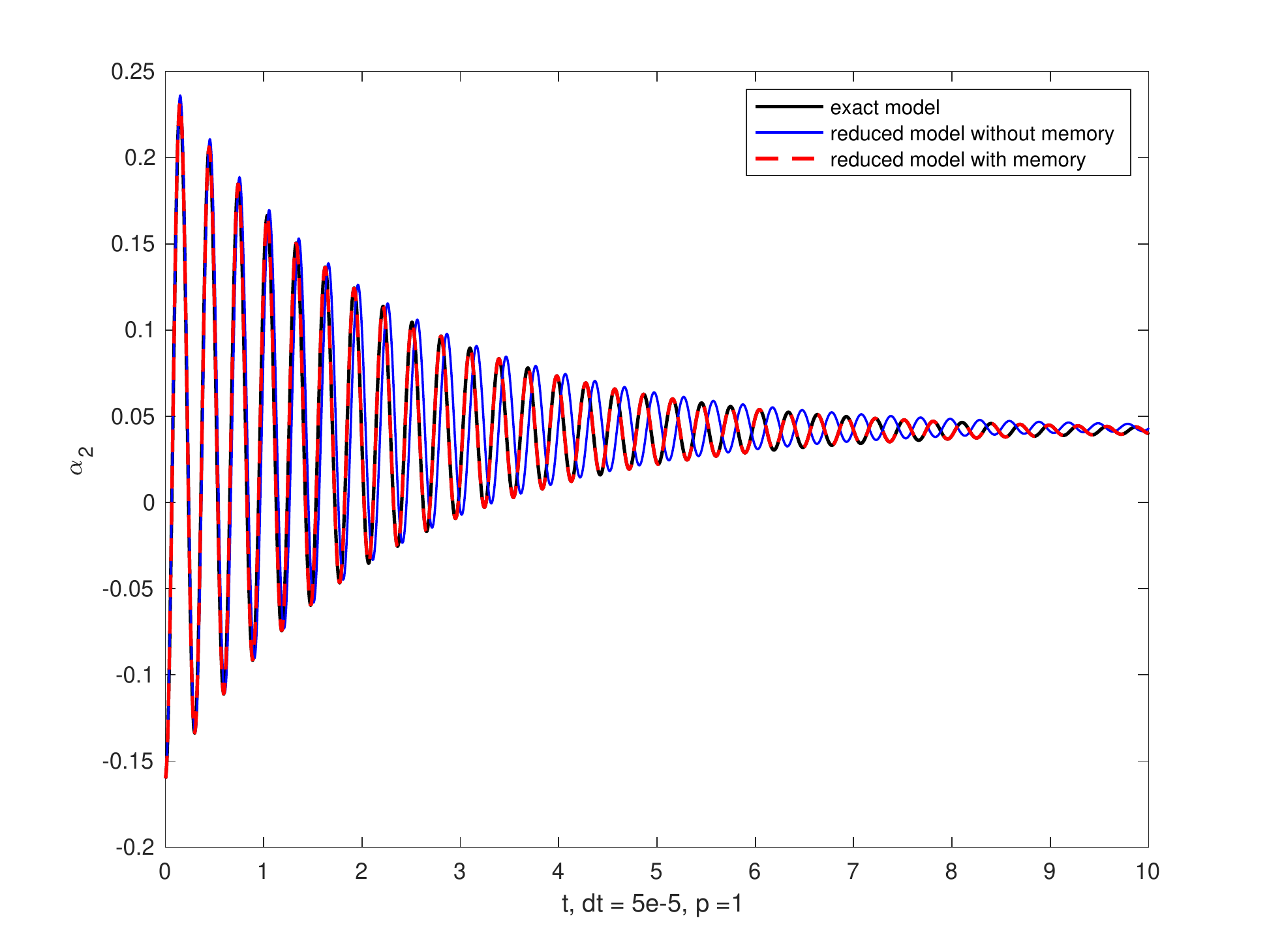}
   \label{fig:infinite1}}
      \quad
   \subfigure[]{%
   \includegraphics[width = 3.5cm]{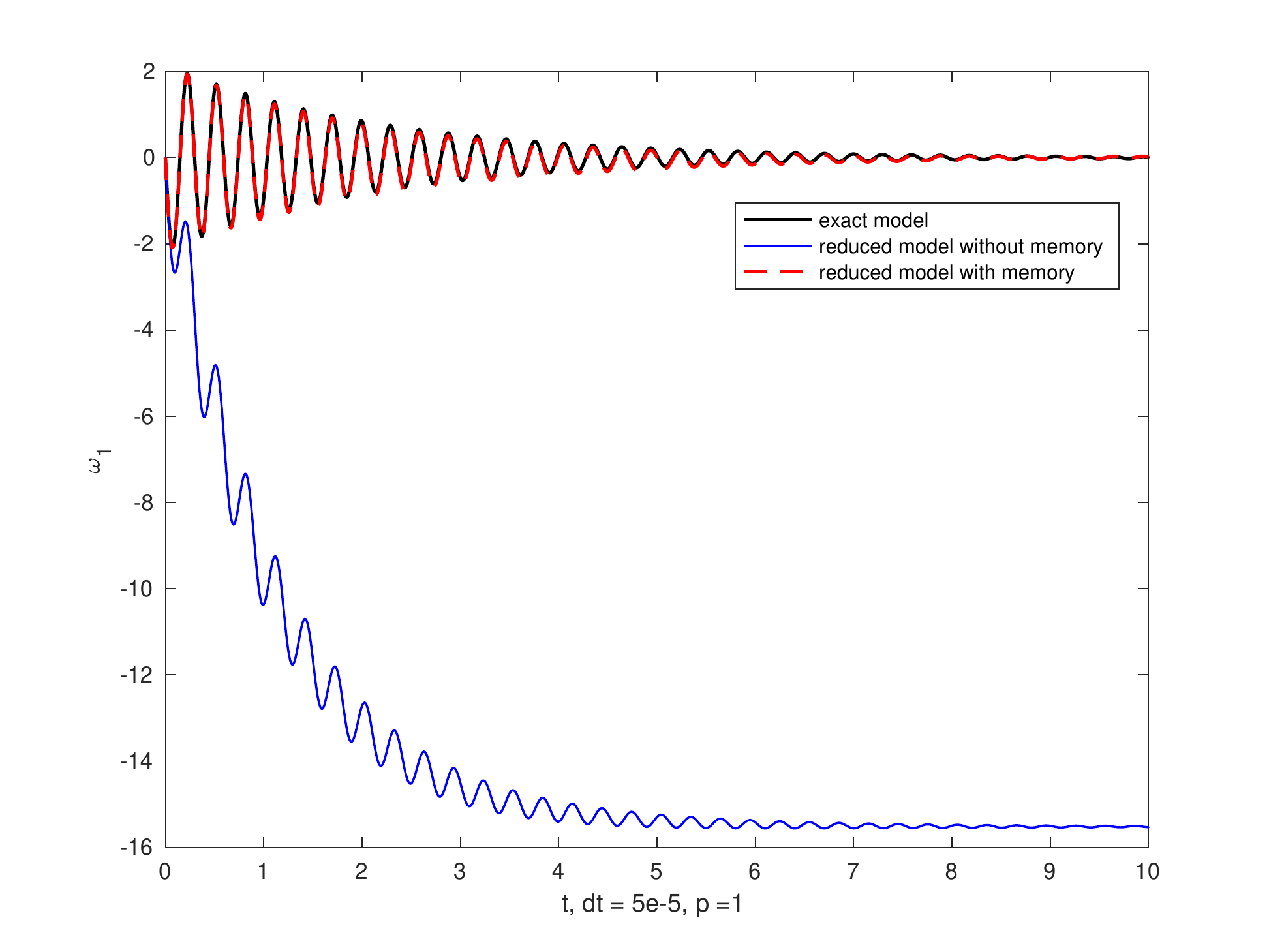}
   \label{fig:infinite2}}
         \quad
   \subfigure[]{%
   \includegraphics[width = 3.5cm]{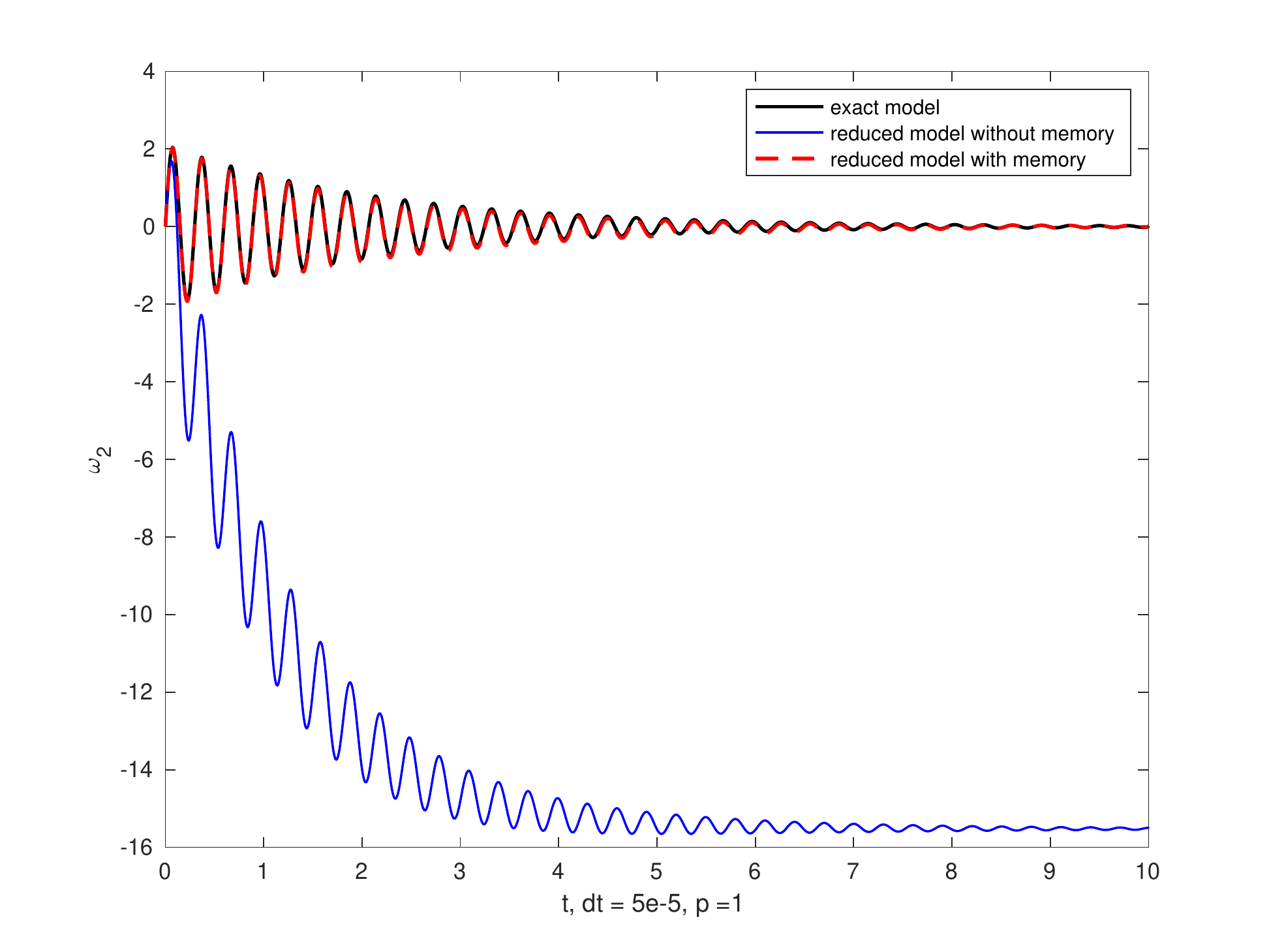}
   \label{fig:infinite3}}
\caption{Reduced model for 3-bus system. Comparison of reduced models with infinite memory and without memory. a) Evolution of the resolved variable $\alpha_2,$ b) Evolution of the resolved variable $\omega_1$ and c) Evolution of the resolved variable $\omega_2.$}
\label{fig:infinite}
\end{figure}

We conducted numerical experiments for several values of the memory length $t_{memory}.$ We present results which show the sensitive dependence of the accuracy of the reduced model on the length of the memory. Figure  \ref{fig:infinite} compares results for the reduced model with infinite memory and the reduced model without memory which includes only the Markovian term in \eqref{eq:redu_b}. We want to make three observations. First, it is obvious that the complete absence of memory is detrimental to the accuracy of the reduced model. Second, that the accuracy degradation rate of the memoryless reduced model is not the same for all the resolved variables. In particular, while the predictions of the memoryless model for $\omega_1$ and $\omega_2$ become highly inaccurate very fast, the prediction for $\alpha_2$ loses the phase but retains the order of magnitude of the exact solution. The reason for this can be easily found in the RHS of equation \eqref{3_bus_rhs3} for the evolution of $\alpha_2.$ The RHS depends {\it only} on the resolved variables $\omega_1$ and $\omega_2.$ Thus, the reduced model equation for this variable does not require the explicit presence of any memory term (see also \eqref{3_bus_rhs3_projected}). Of course, because the other two resolved variables $\omega_1$ and $\omega_2$ are not evolved accurately by the memoryless model, the evolution of $\alpha_2$ is affected, albeit only in terms of losing the phase but not the magnitude of the exact solution.

The third observation concerns the cause of the accuracy degradation of the memoryless model. Note that here we have examined a case where the unresolved variables were not allowed to have any fluctuations. So, the memory effects are not due to the fact that the projection operator we have used also does not allow fluctuations. The memory effects are due to the absence of timescale separation between the resolved and unresolved variables which makes mandatory to account for the history of the resolved variables through a memory term. 

\begin{figure}[htbp]
   \centering
   \subfigure[]{%
   \includegraphics[width = 5cm]{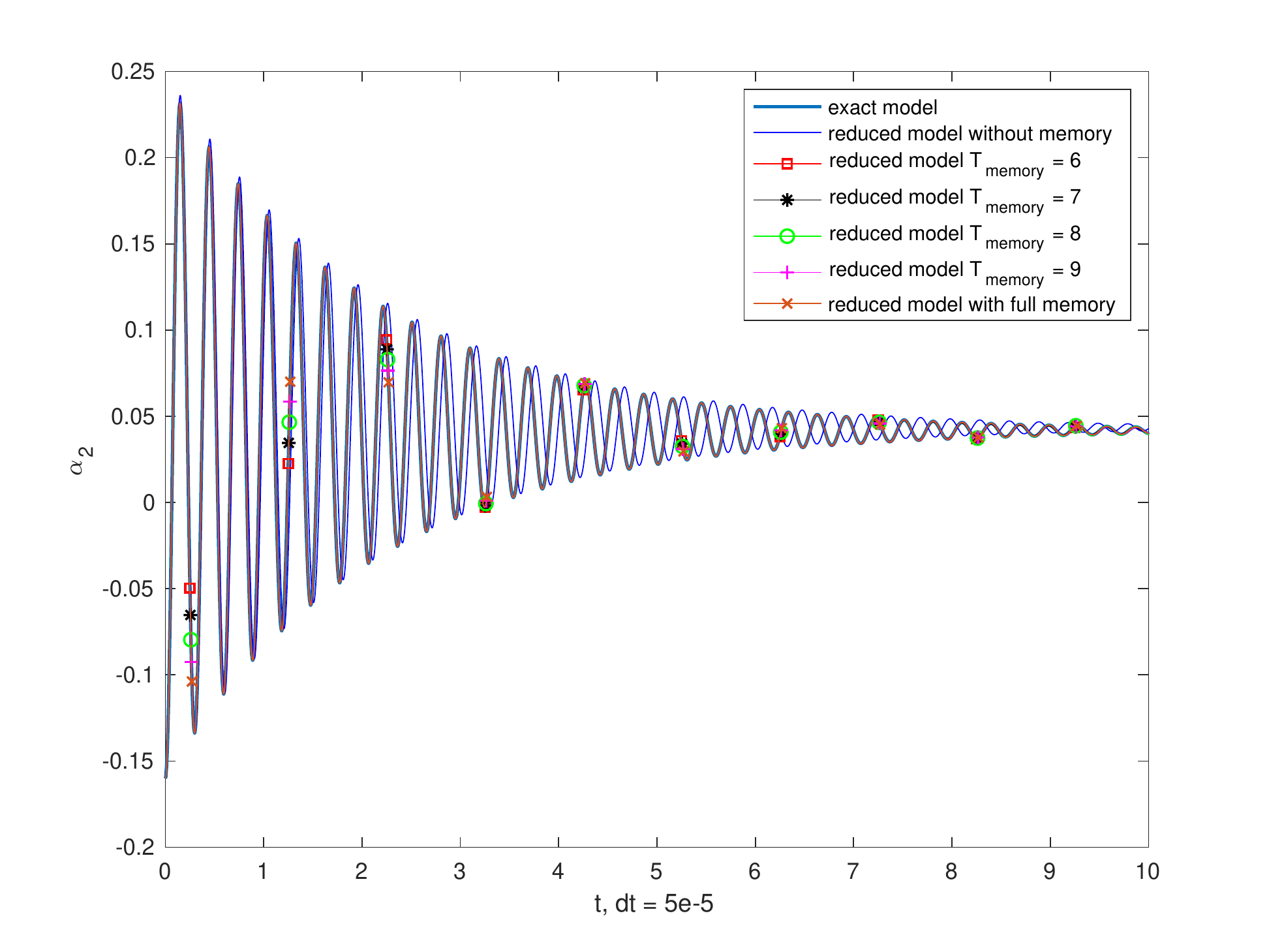}
   \label{fig:finite1}}
      \quad
   \subfigure[]{%
   \includegraphics[width = 5cm]{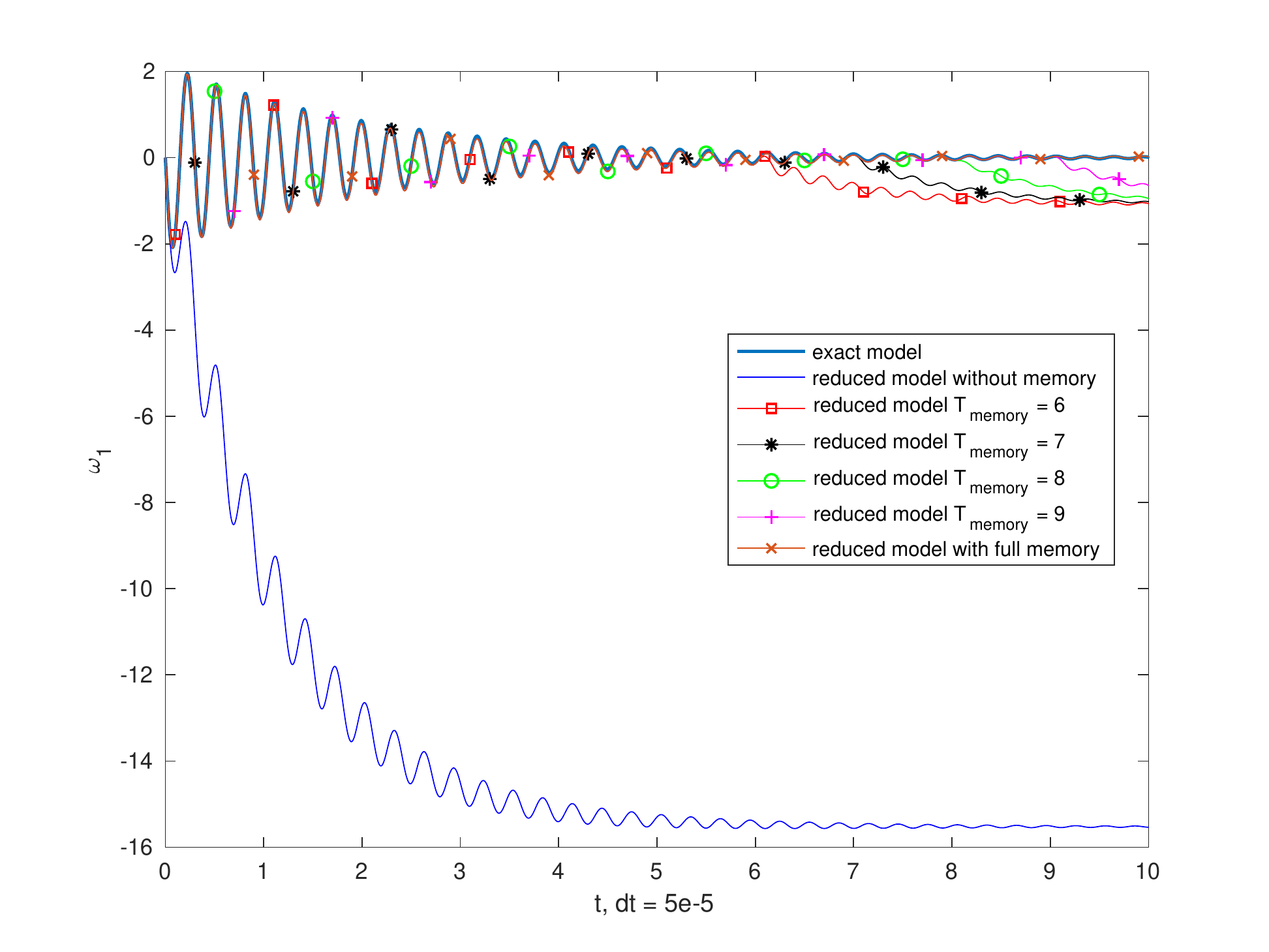}
   \label{fig:finite2}}
\caption{Reduced model for 3-bus system. Comparison of reduced models with variable memory length (including without memory). a) Evolution of the resolved variable $\alpha_2$ and b) Evolution of the resolved variable $\omega_1.$}
\label{fig:finite}
\end{figure}

Motivated by the preceding observation, we investigate the actual length of the history (memory) that is necessary to guarantee accurate predictions of the reduced model for long times. Figure \ref{fig:finite} shows the predictions of the reduced model for various memory lengths $t_{memory}$ (see also \eqref{eq:redu_b_finite}). We want to make two observations. First, the prediction accuracy of the reduced model with memory length $t_{memory}$ starts degrading already for times that are slightly larger than $t_{memory}.$ The second observation is that even if we truncate the memory length, the reduced model does not become unstable. On the contrary, the predictions for $\omega_1$ and $\omega_2$ (not shown) appear to asymptote towards the same long time value.

\begin{figure}[htbp]
   \centering
   \subfigure[]{%
   \includegraphics[width = 5cm]{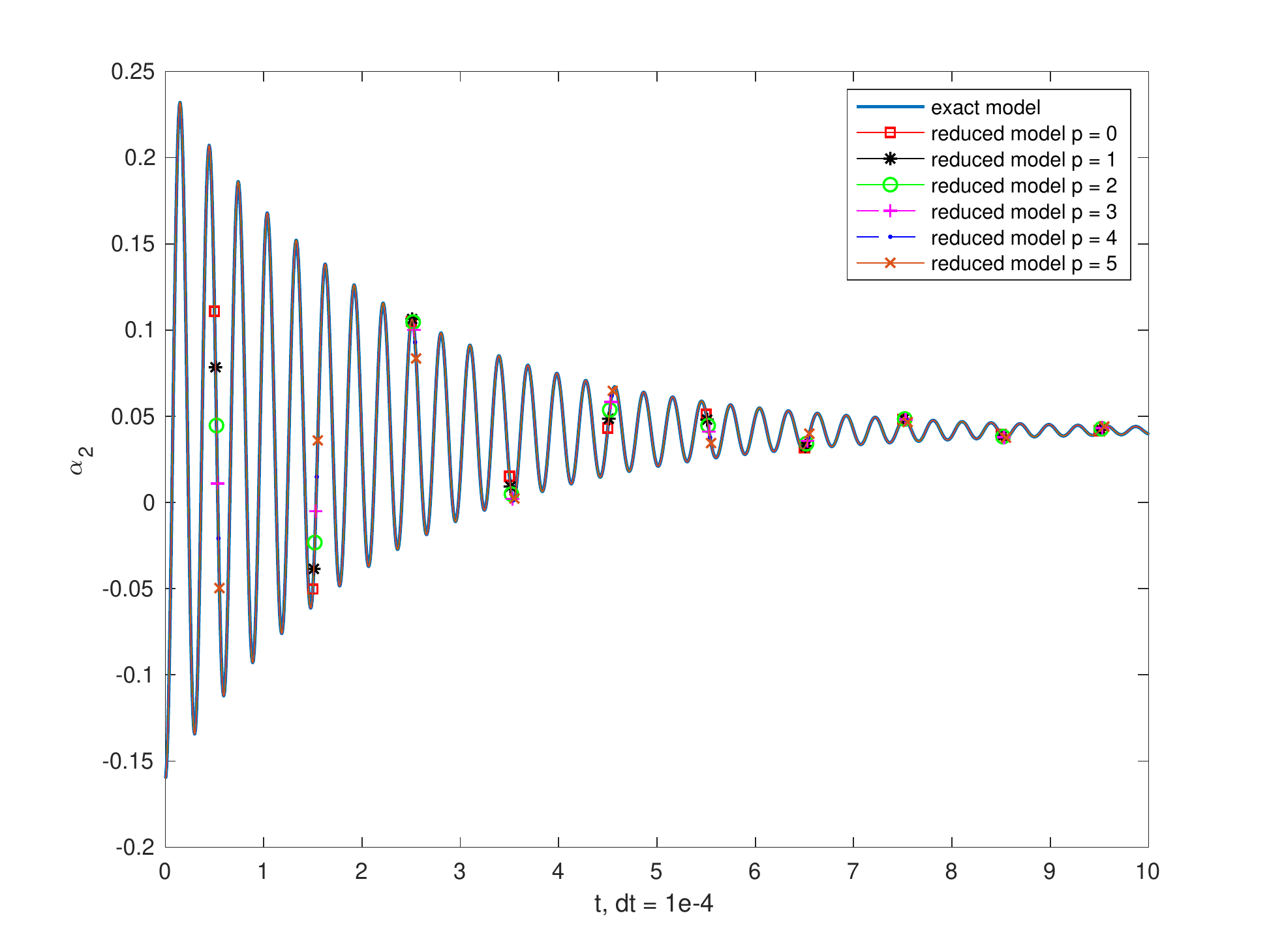}
   \label{fig:finite_longer1}}
      \quad
   \subfigure[]{%
   \includegraphics[width = 5cm]{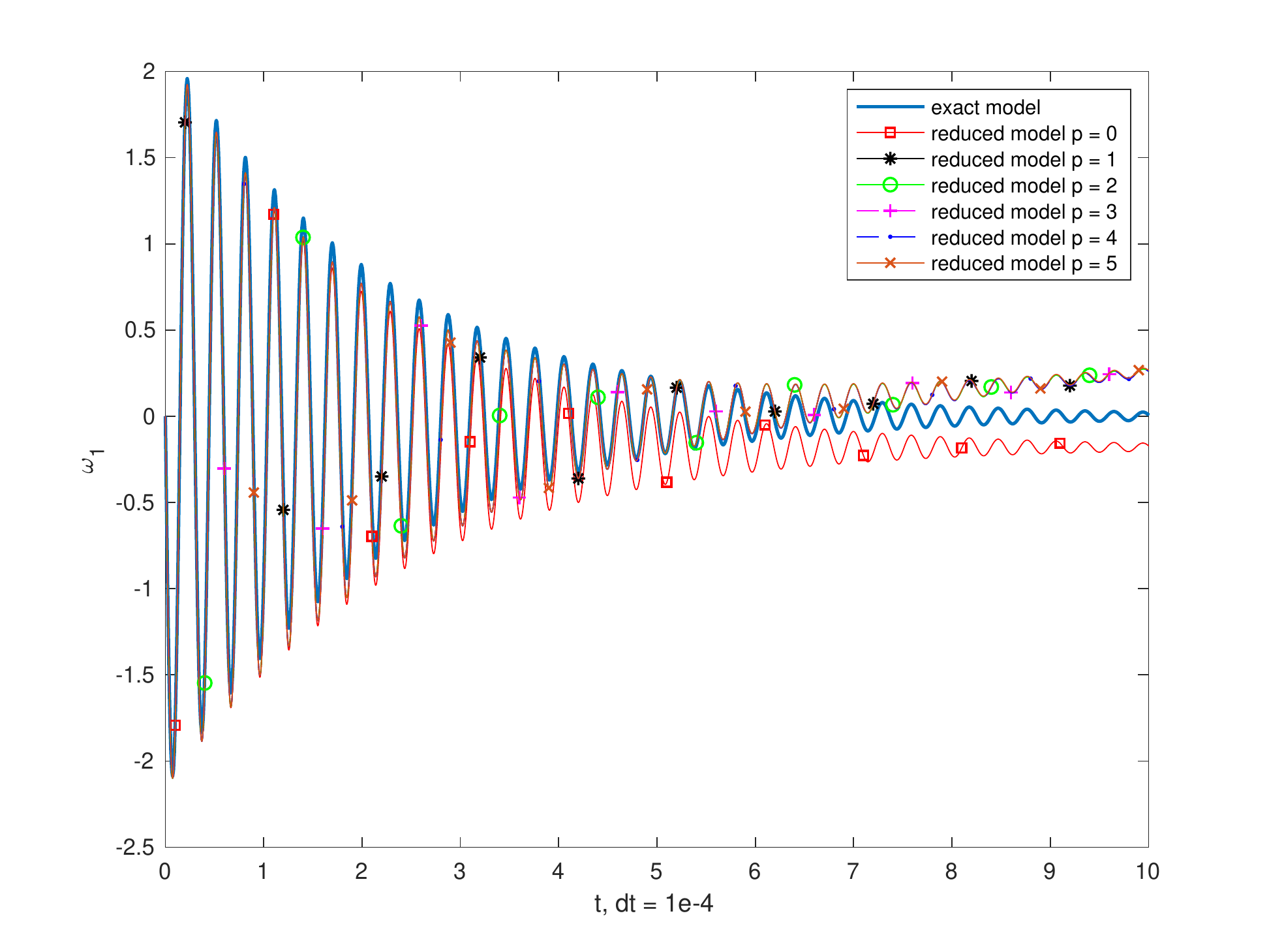}
   \label{fig:finite_longer2}}
\caption{Reduced model for 3-bus system. Comparison of reduced models with variable memory length (including without memory), timestep $\Delta t =10^{-4}$ and $p=5.$ a) Evolution of the resolved variable $\alpha_2,$ b) Evolution of the resolved variable $\omega_1.$}
\label{fig:finite_longer}
\end{figure}

The high accuracy of the reduced model for $\Delta t = 5 \times 10^{-5}$ raises the question of whether we can increase the timestep without suffering a severe accuracy degradation. To this purpose, we conducted numerical experiments with $\Delta t = 10^{-4}$ and we allowed the order $p$ of the finite-rank projection of the memory to vary between 0 and 5. In other words, we want to see if we can compensate for a larger timestep with a higher order finite-rank projection. Figure \ref{fig:finite_longer} shows the evolution of the resolved variables $\alpha_2$ and $\omega_1$ (the evolution of $\omega_2$ is similar and we omit it). The results reveal that while the $p=0$ finite-rank projection of the memory slightly underestimates the solution of the exact model, all the higher order approximations of the memory overestimate it and in fact, become unstable for long times. This means that at least for this example, increasing the timestep cannot be compensated by a higher order finite-rank projection for the memory.  

\subsubsection{The linear representation of the memory}\label{linear_memory}

We make one final remark about the representation of the memory. The results in Figs. \ref{fig:infinite} and \ref{fig:finite} for $\Delta t=5 \times 10^{-5},$ as well as those in Fig. \ref{fig:finite_longer} for $\Delta t =10^{-4}$ show, that the memory can be essentially captured by the {\it linear} basis functions in the finite-rank projection onto the resolved variables, i.e. the functions of order $p=1.$ The higher order basis functions do not seem to contribute substantially to the representation of the memory. We can provide a qualitative explanation of this occurrence by examining closer the evolution of the resolved and unresolved variables in the full system as well as the structure of the RHS of the equations of the DeMarco model. We will focus on the case of the memory of the resolved variable $\omega_1.$ A similar analysis can be done for $\omega_2.$ Finally, note that the reduced equation for $\alpha_2$ has {\it no} memory since the RHS depends only on the resolved variables $\omega_1$ and $\omega_2.$

We begin with three observations: i) from Figs. \ref{fig:a2} and \ref{fig:a3}, we see that the resolved variable $\alpha_2$ and the unresolved variable $\alpha_3$ evolve at very similar timescales and also at similar magnitudes (they are also in phase due to initial conditions); ii) from Fig. \ref{fig:V3}, we see that the unresolved variable $V_3$ does not change much its value during the time interval of simulation and remains close to 1, so its rate of change given by $R_5$ in \eqref{3_bus_rhs5} is small; also from Fig. \ref{fig:a3}, we see that the unresolved variable $\alpha_3$ changes rapidly so its rate of change given by $R_4$ in \eqref{3_bus_rhs4} is large; and iii) from the definition of the parameters in our example and the definition of the projection operator, we see that $F_1(u_0,0)=QL\omega_{10}=a_{1,4} b_2 V_1 [V_{30} \sin(\alpha_{30})-V_3^0\sin(\alpha^0_{3})].$ 

We remind the reader that the memory kernels $b_j^{\nu}(s)=(LF_j(u_0,s),h^{\nu}(\hat{u}_0))$ in \eqref{b_nu_def} require the orthogonal dynamics to compute $F_j(u_0,s),$ which in turn leads to the solution of Volterra equation system in \eqref{Volterra_a}. However, for $s=0$ we can perform some analysis to get some insight about the behavior of the memory kernels $b_j^{\nu}(s).$ We can acquire additional insight about $b_j^{\nu}(s)$ from the quantities $f_j^{\mu}(s)=(Le^{tL}F_j(u_0,0),h^{\mu}(\hat{u}_0))$ which also appear in the Volterra equation \eqref{Volterra_a} and for which we can obtain an analytical expression using the identity in \eqref{memory_22}. The quantities $f_j^{\mu}(s)$ are the analogs of the memory kernels but using the full dynamics and not just the orthogonal dynamics (see \cite{CHK3,CS06} for a more extended discussion).

To proceed, we need to estimate the action of $L$ on $F_1(u_0,0).$ From observation iii) we have that 
\begin{gather}
LF_1(u_0,0)=a_{1,4} b_2 V_1 LV_{30} \sin(\alpha_{30}) + a_{1,4} b_2 V_1V_{30} L\sin(\alpha_{30}) \notag \\
= a_{1,4} b_2 V_1 R_5(u_0)  \sin(\alpha_{30}) +  a_{1,4} b_2 V_1V_{30} R_4(u_0) \cos(\alpha_{30}), 
\end{gather}
where $R_5(u_0)$ and $R_4(u_0)$ are given by \eqref{3_bus_rhs5} and \eqref{3_bus_rhs4} respectively. From observation ii),  the expression $R_5(u_0)$ is small while $R_4(u_0)$ is large. So, the expression $a_{1,4} b_2 V_1 R_5(u_0)  \sin(\alpha_{30})$ is small compared to $a_{1,4} b_2 V_1V_{30} R_4(u_0) \cos(\alpha_{30})$ (recall that $V_{30}$ is close to 1). From this we find that the expression $LF_1(u_0,0) \approx a_{1,4} b_2 V_1V_{30} R_4(u_0) \cos(\alpha_{30}).$ Using \eqref{3_bus_rhs4} for $R_4(u_0)$ we find
\begin{gather}
LF_1(u_0,0) \approx a_{1,4} b_2 V_1V_{30} \bigl[ a_{4,1}M_1\w_{10} \notag \\
+a_{4,4}[b_2V_1V_{30}\sin(\alpha_{30}) + b_3V_2V_{30}\sin(\alpha_{30}-\alpha_{20})+P_3] \bigr]\cos(\alpha_{30}).
\end{gather}
From observation i) we see that $\alpha_3-\alpha_2$ is small and thus, $\sin(\alpha_3-\alpha_2) \approx \alpha_3-\alpha_2.$ So, we have for $LF_1(u_0,0)$ that  
\begin{gather}
LF_1(u_0,0) \approx a_{1,4} b_2 V_1V_{30} \bigl[ a_{4,1}M_1\w_{10} \notag \\
+a_{4,4}[b_2V_1V_{30}\sin(\alpha_{30}) + b_3V_2V_{30}(\alpha_{30}-\alpha_{20})+P_3] \bigr]\cos(\alpha_{30}) \label{linear_final}.
\end{gather}
From \eqref{linear_final} we see that $LF_1(u_0,0)$ can be reasonably approximated by a {\it linear} function of the resolved variables $\omega_{10}$ and $\alpha_{20}.$ Finally, if we inspect Figs. \ref{fig:w1} and \ref{fig:w2} we see that $\omega_1$ and $\omega_2$ are in phase, of comparable magnitude and opposite signs. So, $LF_1(u_0,0)$ will depend also approximately linearly on $\omega_{20}$ too. From these considerations, we expect that the finite-rank projection of the memory for $\omega_1$ on the resolved variables will essentially contain only {\it linear} basis functions. 

\begin{figure}[htbp]
   \centering
   \subfigure[]{%
   \includegraphics[width = 3.5cm]{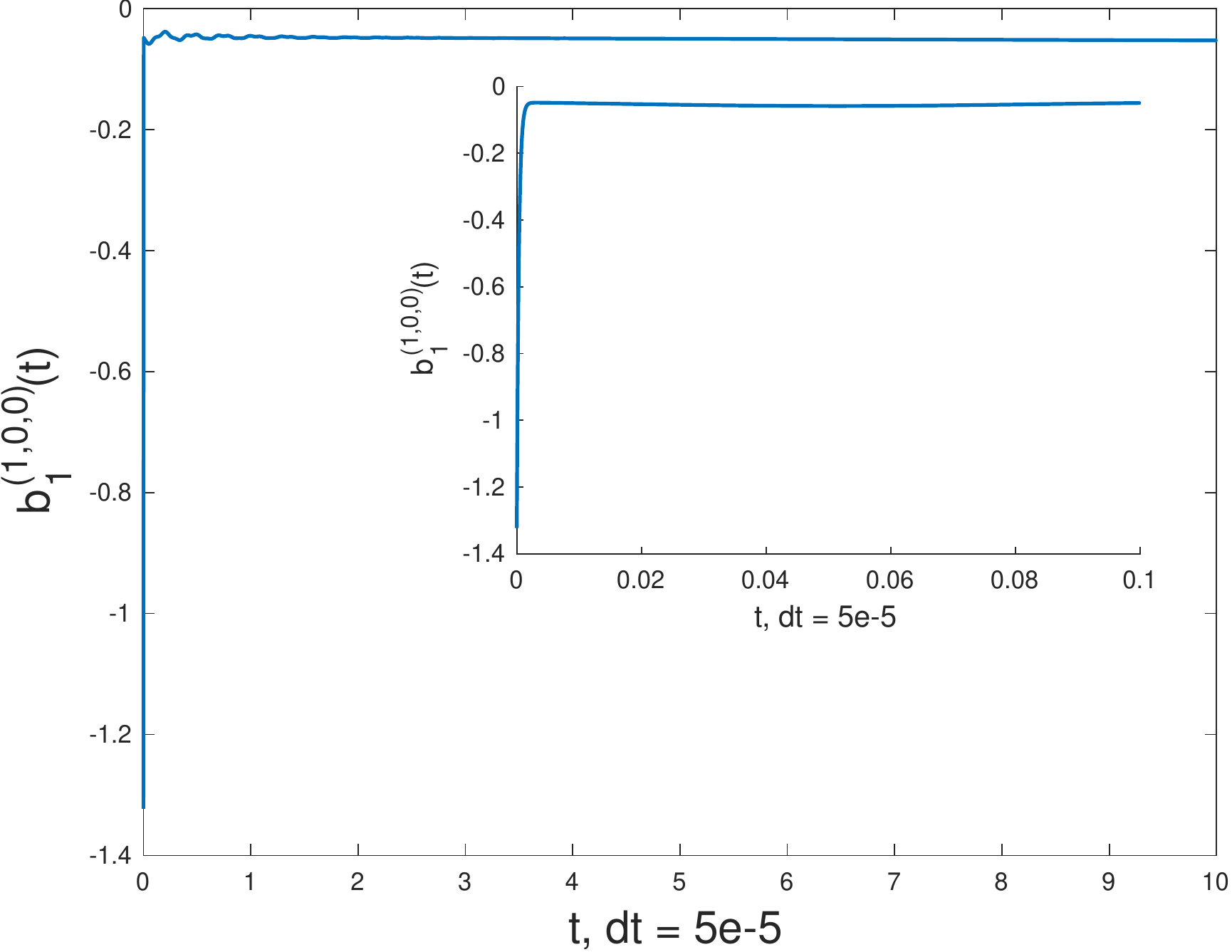}
   \label{fig:linear1}}
      \quad
   \subfigure[]{%
   \includegraphics[width = 3.5cm]{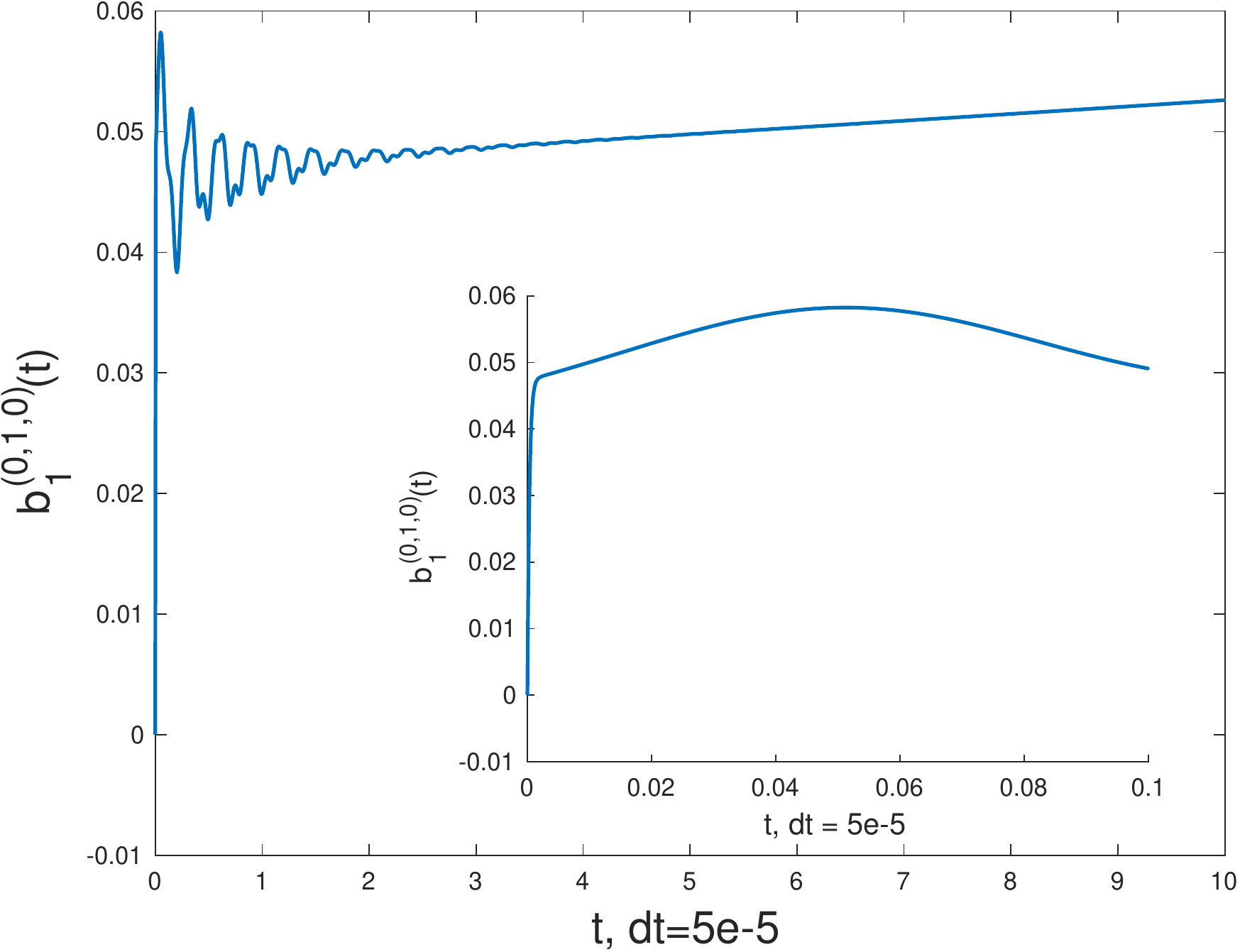}
   \label{fig:linear2}}
      \quad
   \subfigure[]{%
   \includegraphics[width = 3.5cm]{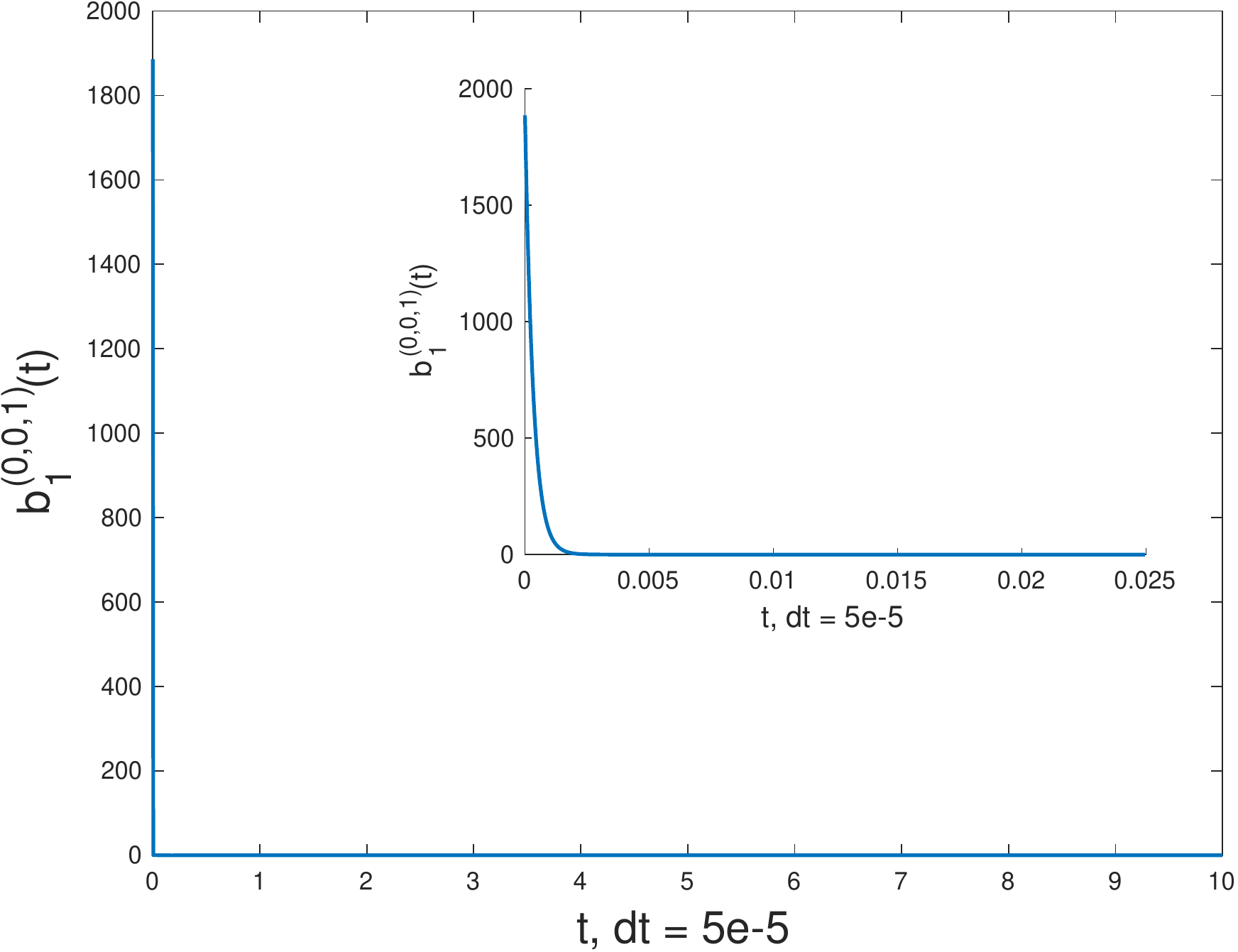}
   \label{fig:linear3}}   
\caption{Reduced model for 3-bus system. Evolution of the memory kernels $b_1^{\mu}(s)$ of the resolved variable $\omega_1$ on linear functions of the resolved variables. a) Projection on the 1 degree Hermite polynomial $h^{(1,0,0)},$ b) Projection on the 1 degree Hermite polynomial $h^{(0,1,0)},$ and c) Projection on the 1 degree Hermite polynomial $h^{(0,0,1)}.$ We have also included in the insets the evolution near the time origin (see text for details).}
\label{fig:linear}
\end{figure}

Figs. \ref{fig:linear1}-\ref{fig:linear3} corroborate our brief analysis. Fig. \ref{fig:linear1} shows the evolution of the memory kernel $b_1^{(1,0,0)}(s)$ which is the projection coefficient of $LF_1(u_0,s)$ on the degree 1 Hermite polynomial $h^{(1,0,0)}(\hat{u}_0)=2\omega_{10}.$ The memory kernel $b_1^{(1,0,0)}(s)$ decays quickly to the asymptotic value -0.0526. The memory kernel $b_1^{(0,1,0)}(s)$ shown in Fig. \ref{fig:linear2} is the projection coefficient of $LF_1(u_0,s)$ on the degree 1 Hermite polynomial $h^{(0,1,0)}(\hat{u}_0)=2\omega_{20}.$ It asymptotes to the value 0.0526. This is not accidental since as we have mentioned above $\omega_1$ and $\omega_2$ are related. Finally, Fig. \ref{fig:linear3} shows the memory kernel $b_1^{(0,0,1)}(s)$ which is the projection coefficient of $LF_1(u_0,s)$ on the degree 1 Hermite polynomial $h^{(0,0,1)}(\hat{u}_0)=2\alpha_{20}.$ It decays quickly to the value -0.0048. Again this is to be expected because the dependence of $LF_1(u_0,0)$ on $\alpha_{20}$ is weak since it only enters through the difference with $\alpha_{30}$ which has similar magnitude.

\begin{figure}[htbp]
   \centering
   \subfigure[]{%
   \includegraphics[width = 3.5cm]{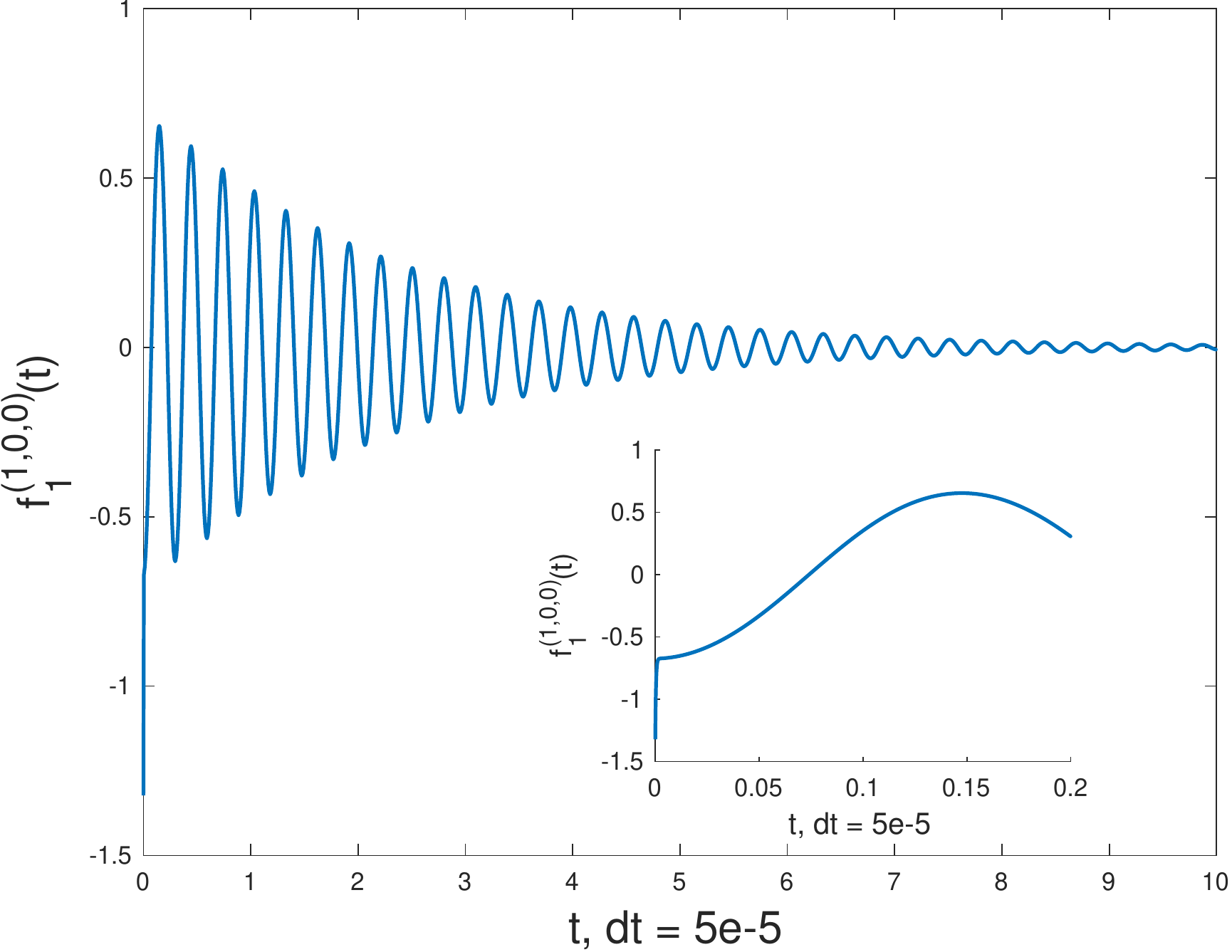}
   \label{fig:full_linear1}}
      \quad
   \subfigure[]{%
   \includegraphics[width = 3.5cm]{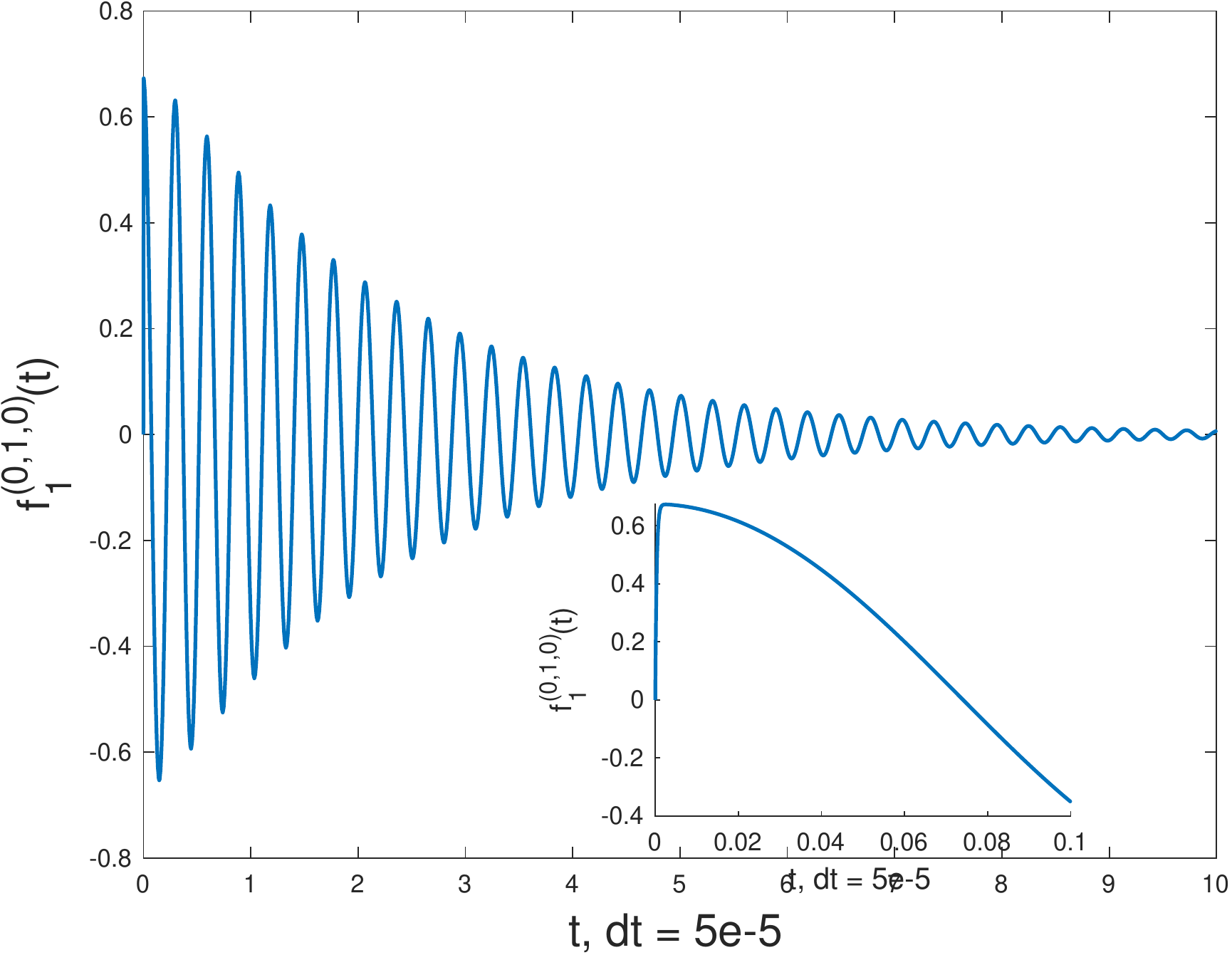}
   \label{fig:full_linear2}}
      \quad
   \subfigure[]{%
   \includegraphics[width = 3.5cm]{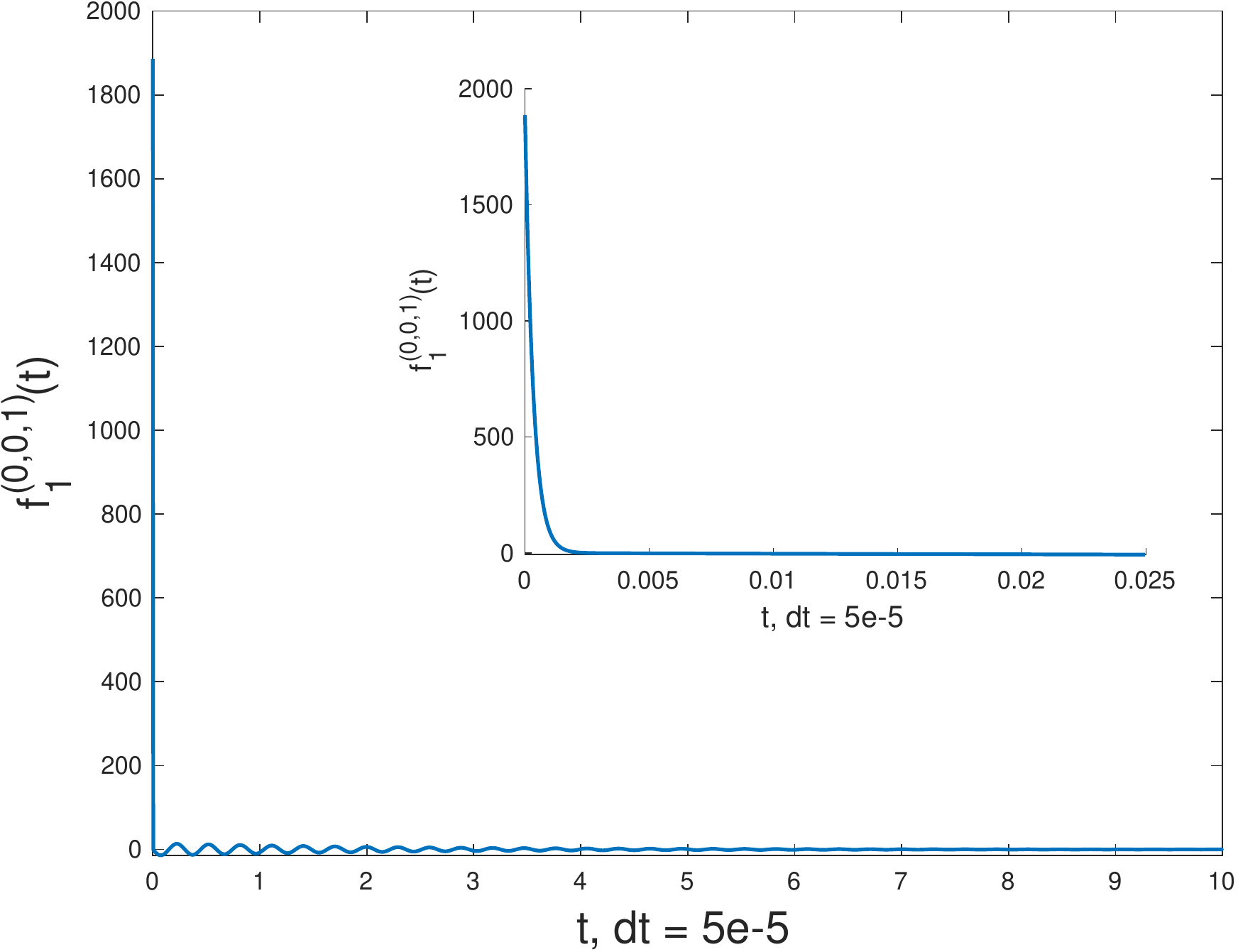}
   \label{fig:full_linear3}}   
\caption{Reduced model for 3-bus system. Evolution of the projections $f_1^{\mu}(s)=(Le^{sL}F_1(u_0,0),h^{\mu}(\hat{u}_0))$ of $Le^{sL}F_1(u_0,0)$ on linear functions of the resolved variables.  a) Projection on the 1 degree Hermite polynomial $h^{(1,0,0)},$ b) Projection on the 1 degree Hermite polynomial $h^{(0,1,0)},$ and c) Projection on the 1 degree Hermite polynomial $h^{(0,0,1)}.$ We have also included in the insets the evolution near the time origin (see text for details).}
\label{fig:full_linear}
\end{figure}

Figs. \ref{fig:full_linear1}-\ref{fig:full_linear3} show the evolution of $f_1^{\mu}(s)=(Le^{sL}F_1(u_0,0),h^{\mu}(\hat{u}_0))$ which are used to setup the Volterra equation \eqref{Volterra_a}. As we have mentioned, these quantities are the analogs of the memory kernels $b_1^{\mu}(s)$ but they are computed using the full dynamics instead of the orthogonal dynamics. Comparing Figs.  \ref{fig:linear1}-\ref{fig:linear3} with Figs.  \ref{fig:full_linear1}-\ref{fig:full_linear3}, we see that even though $b_1^{\mu}(s)$ and $f_1^{\mu}(s)$ are similar for small $s$ (they have to be due to their definitions), they start deviating fast from one another. Even though the $f_1^{\mu}(s)$ follow decaying oscillations, the memory kernels $b_1^{\mu}(s)$ settle relatively fast to a small but {\it nonzero} value. This nonzero value is the reason that we need to keep track of the history of the resolved variables for so long in the distant past. We note that such a behavior is peculiar for systems with long memory where one expects the memory kernels to keep oscillating for long times (see e.g. the analytically tractable example in Section \ref{exact_linear_oscillator}). It must be related to the special structure of the DeMarco model and we plan to investigate more in a future publication. 

In summary, despite the peculiarity of the behavior of the memory kernels, our qualitative analysis shows that the adequate representation of the memory using only linear basis functions is not surprising. It will be interesting to see what is the behavior of the memory kernels for a DeMarco model with significantly more buses or for even more complicated power grid models.

\subsubsection{Exact reduced model for a linear oscillator system}\label{exact_linear_oscillator}
Motivated by the pattern we observed for the behavior of the reduced model, we study in this section results for the reduced model of a single particle coupled linearly to a harmonic oscillator heat bath \cite{zwanzig2001}. This system is much simpler than the DeMarco model but our motivation is twofold: i) examine a system with very long memory where an exact reduced model for a part of it can be derived analytically and ii) show that the accuracy of the exact reduced model depends crucially on the length of the retained history in the memory term. Through this we want to show that, given the approximations we employed to compute the reduced model and the complexity of the DeMarco model, the results we obtained for the reduced model are rather encouraging, and, in a sense, optimal. 

They are optimal, in the sense that the same qualitative behavior exhibited for the reduced model of the DeMarco model appears also in the case of the exact reduced model of a single particle coupled linearly to a heat bath. In particular, if we truncate the memory of the reduced model, then the prediction of the evolution of the single particle loses accuracy fast for times longer than the memory length. In addition, the truncated memory model loses only accuracy but not stability. Finally, even the inclusion of a short memory is much better than not to include any memory at all.   

The particle is described by a coordinate $x$ and its conjugate momentum
$p.$ The heat bath is described by a set of coordinates $q_j$ and their
conjugate momenta $p_j$ For simplicity, all oscillator masses are set
equal to 1. The particle Hamiltonian $H_s$ is 
\begin{equation}\label{linear_1}
H_s=\frac{p^2}{2m} + U(x)
\end{equation} 
where $U(x)$ is a potential. We have taken $U(x)=\cos(2x).$ The heat bath Hamiltonian $H_B$ is given by 
\begin{equation}\label{linear_2}
H_B = \sum_j \left (  \frac{p_j^2}{2} + \frac{1}{2} \omega_j^2 \left (q_j - \frac{\gamma_j}{\omega_j^2}x \right )^2  \right),
\end{equation}
where $\omega_j$ is the frequency of the $j$th oscillator and $\gamma_j$ measures the
strength of coupling of the particle to the $j$th oscillator. The equations of evolution of the system of the particle and the heat bath are given by
\begin{eqnarray}
\frac{dx}{dt}=\frac{p}{m}, && \frac{dp}{dt}=-U'(x)+\sum_j \gamma_j \left ( q_j - \frac{\gamma_j}{\omega_j^2}\right ) \label{linear_3}\\
\frac{dq_j}{dt}=p_j, && \frac{dp_j}{dt}=-\omega_j^2 q_j + \gamma_j x. \label{linear_4}
\end{eqnarray}
Assuming that $x(t)$ is known, the evolution of each $q_j$ can be found from \eqref{linear_4},
\begin{equation}\label{linear_5}
q_j(t)=q_j(0)\cos(\omega_j t)+p_j(0) \frac{\sin \omega_j t}{\omega_j} +\gamma_j \int_0^t ds x(s) \frac{\sin \omega_j (t-s)}{\omega_j}. 
\end{equation}
Integration by parts in \eqref{linear_5} allows us to obtain an expression which involves $\frac{p(s)}{m}$ instead of $x(s)$ and which will yield a closed equation for $p(t).$ In particular, 
\begin{gather}
q_j(t) - \frac{\gamma_j}{\omega_j^2} x(t)= \left ( q_j (0)- \frac{\gamma_j}{\omega_j^2}x(0) \right )\cos(\omega_j t) + p_j(0) \frac{\sin \omega_j t}{\omega_j} \notag \\
- \gamma_j \int_0^t ds \frac{p(s)}{m} \frac{\cos \omega_j (t-s)}{\omega_j^2}. \label{linear_6}
\end{gather}
We use \eqref{linear_6} in \eqref{linear_3} and we find
\begin{equation}\label{linear_7}
\frac{dp}{dt}=-U'(x) - \int_0^t ds K(s) \frac{p(t-s)}{m} + F_p(t),
\end{equation}
where the memory kernel $K(t)$ is given by
\begin{equation}\label{linear_8}
K(t) = \sum_j \frac{\gamma_j^2}{\omega_j^2}\cos(\omega_j t)
\end{equation}
and the noise $F_p(t)$ is given by 
\begin{equation}\label{linear_9}
F_p(t) = \sum_j \gamma_j p_j(0) \frac{\sin \omega_j t}{\omega_j} + \sum_j \gamma_j \left ( q_j (0)- \frac{\gamma_j}{\omega_j^2}x(0) \right )\cos(\omega_j t).
\end{equation}
Different choices of frequencies $\omega_j$ and coupling constants $\gamma_j$ lead to different behaviors of the memory kernel $K(t)$ and the noise $F_p(t).$ In our numerical simulations the heat bath consists of 5 oscillators. The parameters are $\gamma_j = 1/(j/3+1)$ and $\omega_j=j$ for $j=1,\ldots,5.$ Also, we chose the initial positions $q_j(0)$ and momenta $p_j(0)$ of the heat bath oscillators equal to 1. For the particle, we chose $x(0)=0$ and $p(0)=1,$ and we also set the mass $m=1.$

\begin{figure}[htbp]
   \centering
   \subfigure[]{%
   \includegraphics[width = 3cm]{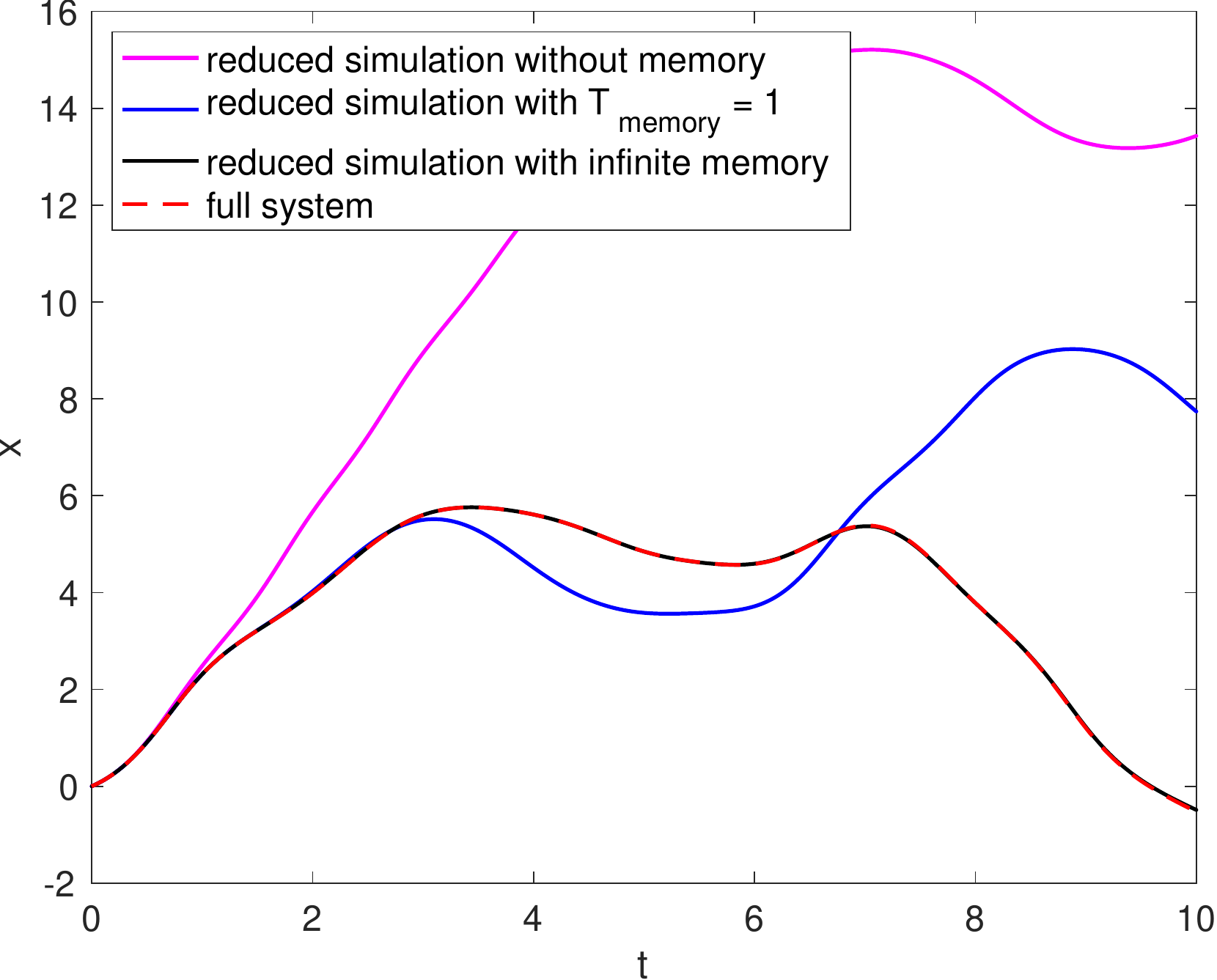}
   \label{fig:xlinear1}}
      \quad
   \subfigure[]{%
   \includegraphics[width = 3cm]{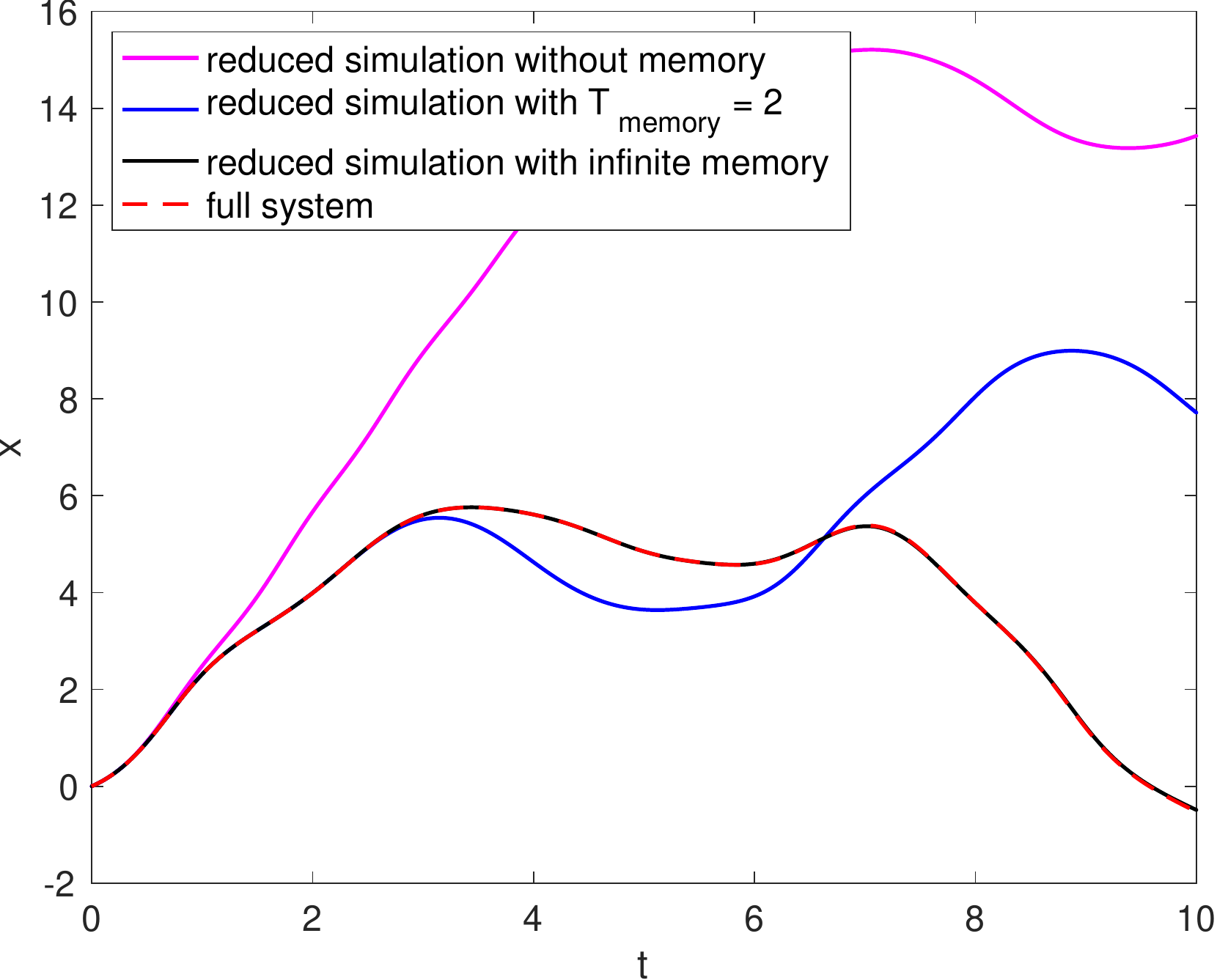}
   \label{fig:xlinear2}}
         \quad
   \subfigure[]{%
   \includegraphics[width = 3cm]{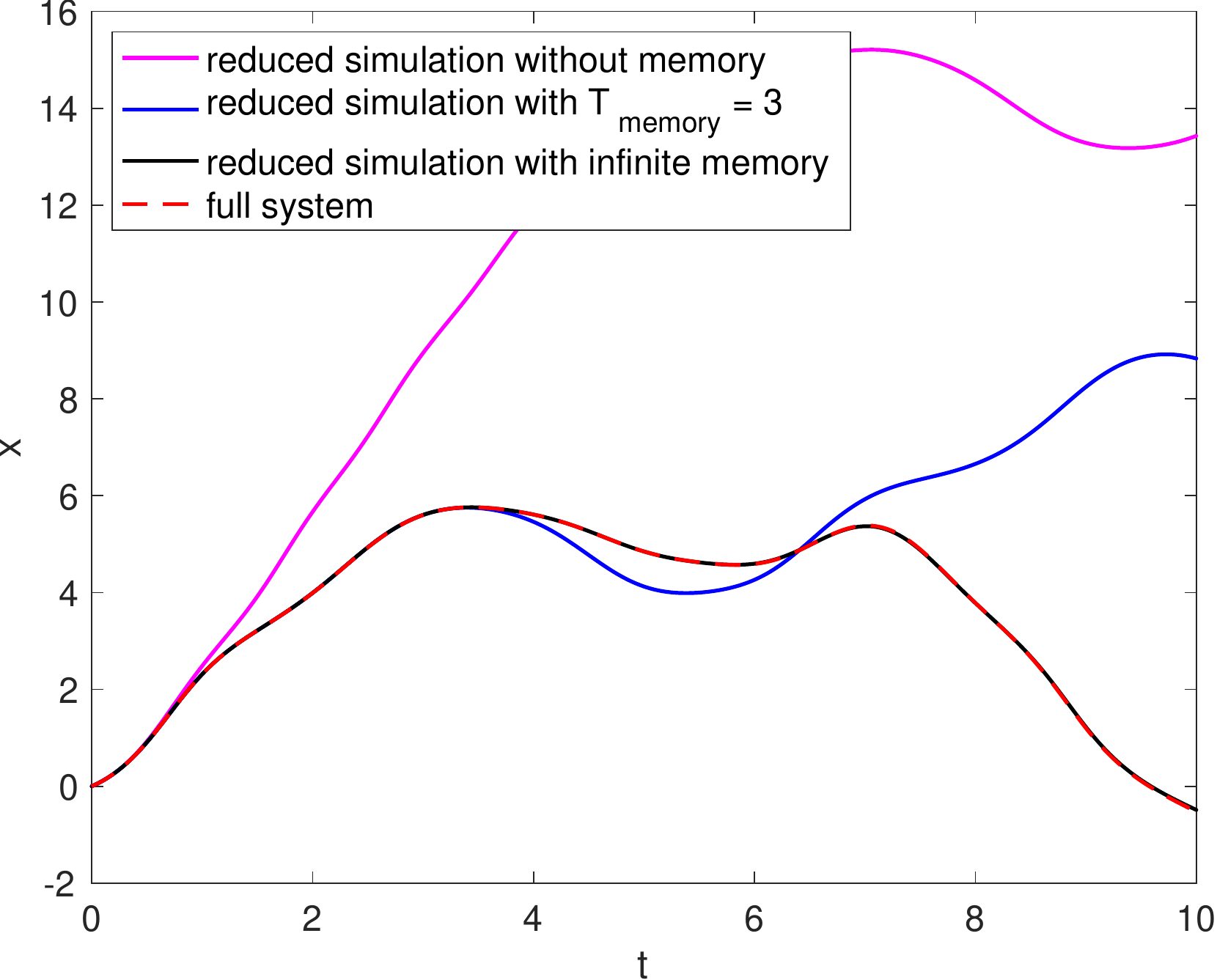}
   \label{fig:xlinear3}}
\caption{Reduced model for particle coupled to heat bath - Evolution of the position $x(t)$ of the particle a) For $t_{memory}=1,$ b) For $t_{memory}=2,$ and c) For $t_{memory}=3.$}
\label{fig:xlinear}
\end{figure}

\begin{figure}[htbp]
   \centering
   \subfigure[]{%
   \includegraphics[width = 3cm]{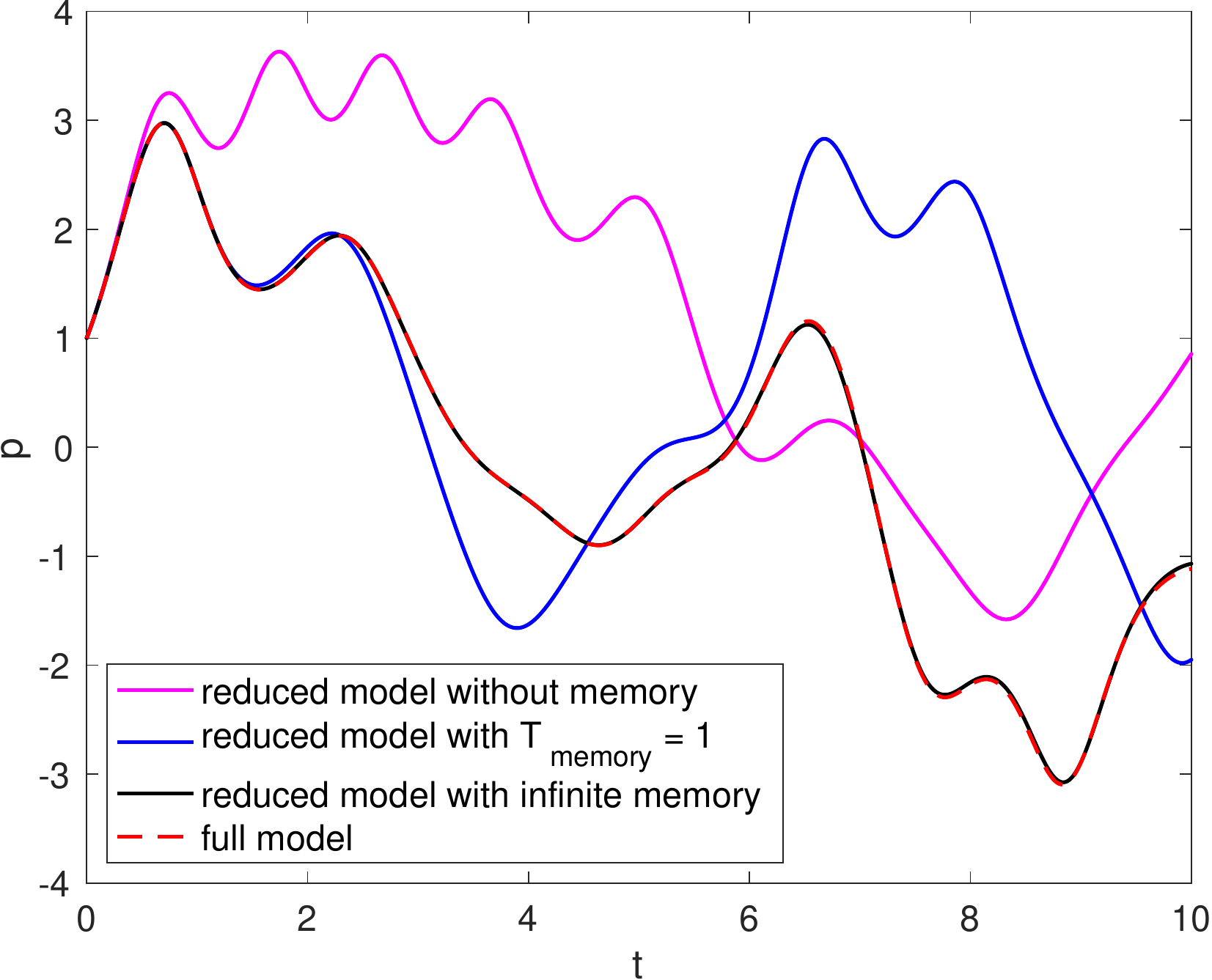}
   \label{fig:plinear1}}
      \quad
   \subfigure[]{%
   \includegraphics[width = 3cm]{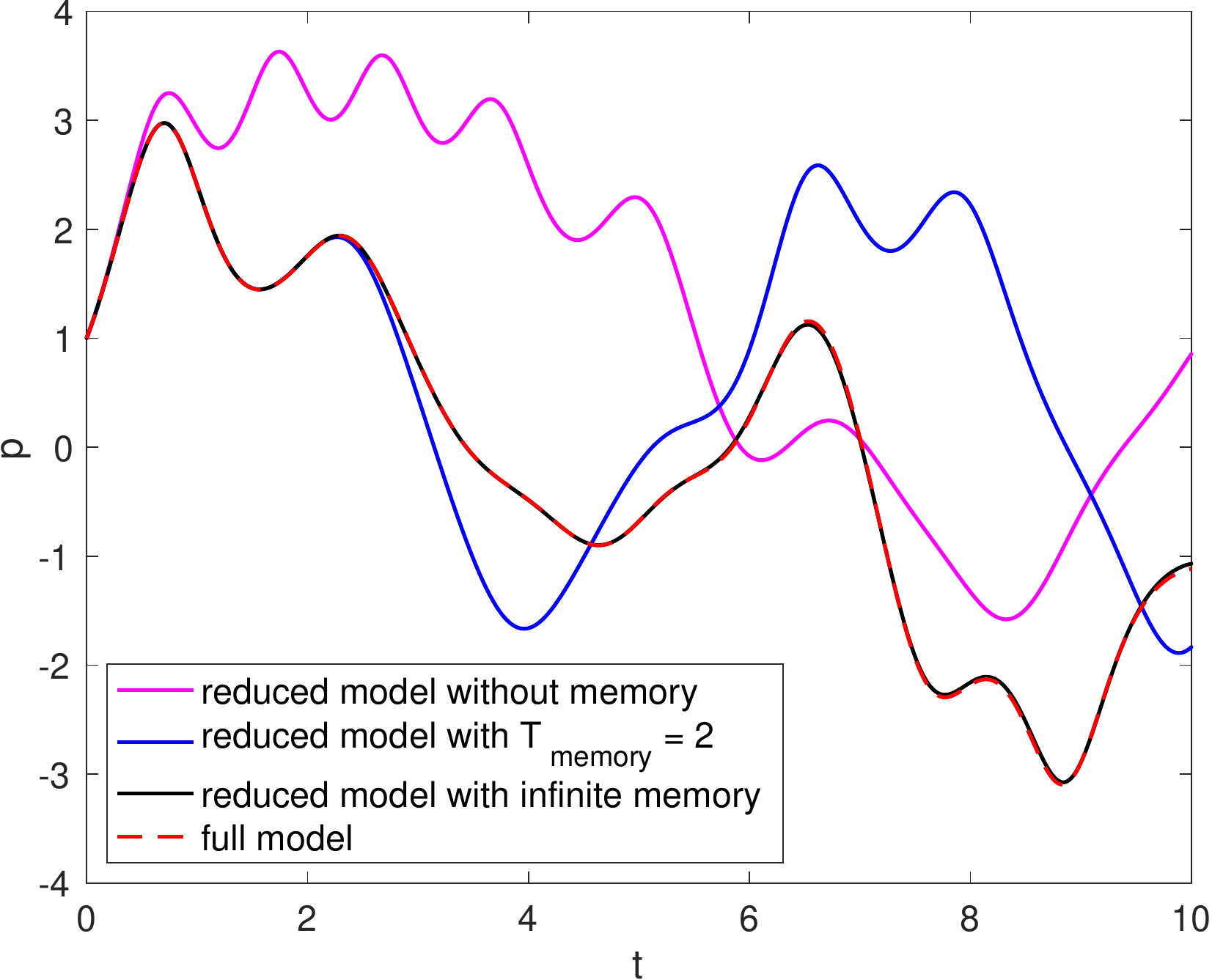}
   \label{fig:plinear2}}
         \quad
   \subfigure[]{%
   \includegraphics[width = 3cm]{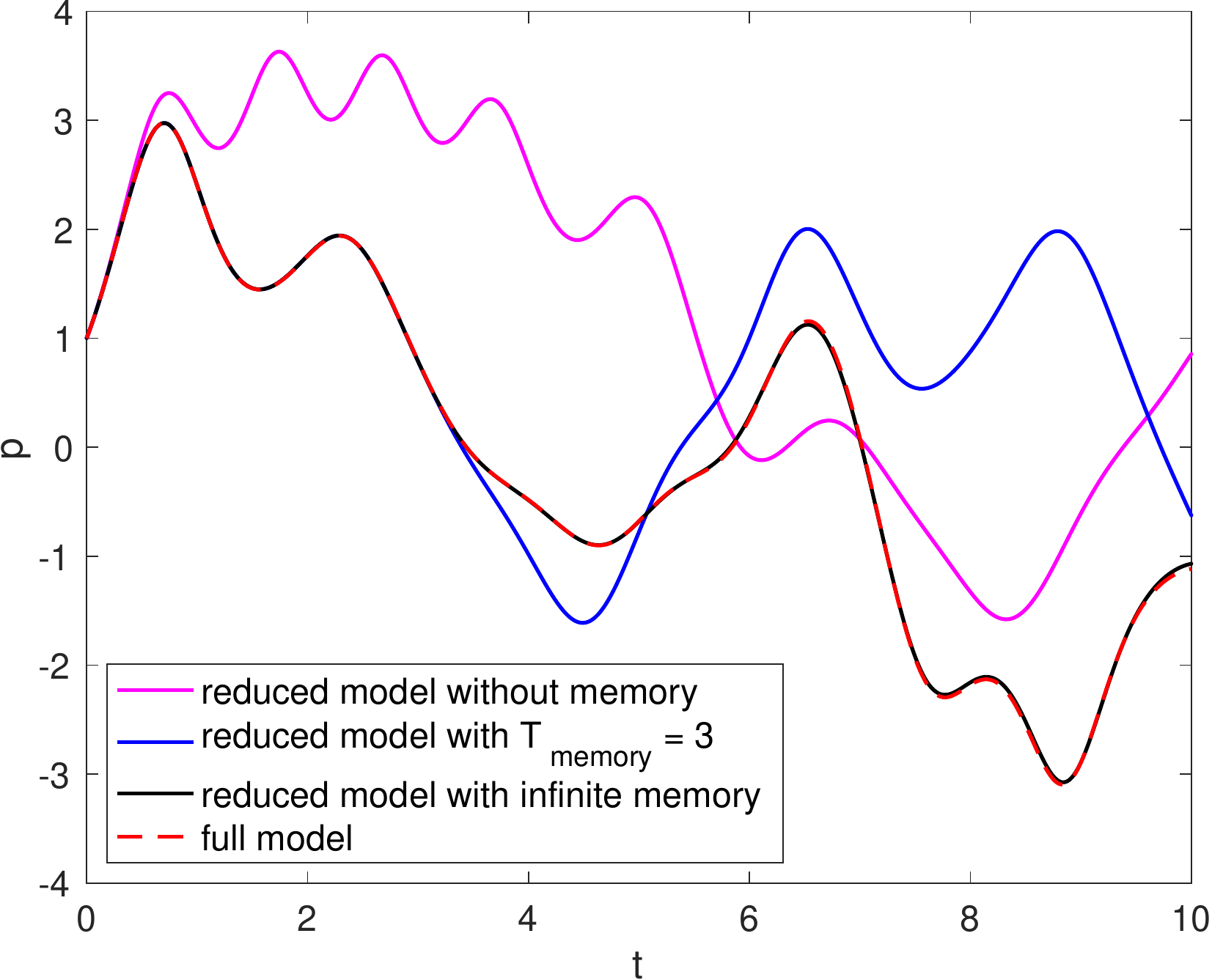}
   \label{fig:plinear3}}
\caption{Reduced model for particle coupled to heat bath - Evolution of the momentum $p(t)$ of the particle a) For $t_{memory}=1,$ b) For $t_{memory}=2,$ and c) For $t_{memory}=3.$}
\label{fig:plinear}
\end{figure}

In Figures \ref{fig:xlinear} and \ref{fig:plinear} we compare the evolution of the particle's position and momentum as predicted by the full system, the exact reduced model \eqref{linear_7}, the reduced model \eqref{linear_7} {\it without} the memory term and reduced models with memory of various lengths $t_{memory}.$  We note that when the memory length is only $t_{memory},$ the reduced model for the particle becomes
\begin{equation}\label{linear_10}
\frac{dp}{dt}=-U'(x) - \int_0^{t_{memory}} ds K(s) \frac{p(t-s)}{m} + F_p(t).
\end{equation}

It is clear from Figs. \ref{fig:xlinear} and \ref{fig:plinear} that the truncation of the memory length affects drastically the accuracy of the reduced model. A reduced model which does not include a memory term loses accuracy very fast. In addition, for reduced models with finite memory $t_{memory},$ the accuracy of the reduced model predictions for both the position and the momentum degrade rapidly for time intervals larger than $t_{memory}.$  These results show that even for simple systems with long memory, extending the temporal interval of accurate predictions of a reduced model is a non-trivial task. Also, it shows that the fast increase of the error for the finite memory reduced model for the DeMarco model for times longer than $t_{memory},$ is not due to the inadequacy of the finite-rank projection of the memory but rather to the failure to include all the necessary history of the resolved variables.


\section{Discussion and future work}\label{discussion}
We have presented results from the Mori-Zwanzig formalism for the construction of reduced order models for the DeMarco power grid model. Even though the DeMarco model can be considered an idealization, it exhibits important qualitative features of more realistic models. In particular, there is absence of timescale separation between generators and load buses which can complicate the construction of an accurate reduced order model for subsets of the state variables. 

Our results corroborate the expectation that in systems with absence of timescale separation between resolved and unresolved variables, it is imperative to account for long memory effects in order to construct an accurate reduced order model. Truncating the memory length can lead to loss of accuracy for integration times that are longer than the memory length. However, even the inclusion of a short memory in the reduced order model results in significant improvement over a memoryless reduced model. This is an important result because in power grid applications, one may be more interested in short time dynamics e.g. for planning purposes, where one could benefit by having a reduced order model. In such scenarios, there could exist the impression that short time predictions with a memoryless model can have acceptable accuracy. Our results present strong evidence that this is not the case and memory has to be included. There is a silver lining to the presented results, namely that, at least for the model investigated, truncating the length of the memory may lead to loss of accuracy but stability is preserved. This is also important, given the fact that stability is a major concern about reduced order models.

An interesting research direction which is relevant to power grid applications concerns allowing the existence of fluctuations for the initial conditions of the unresolved variables. As we have explained in Section \ref{mz_formalism_projection}, our choice of projection operator commutes with a nonlinear function exactly because it does not allow any fluctuations in the unresolved variables. This makes Eq. \eqref{mzp} of the MZ formalism valid {\it pathwise}. However, as we saw in Section \ref{3_bus_reduced_memory}, in order to obtain an analytical expression for the memory kernel we had to approximate our projection operator by a finite-rank projection operator. The finite-rank projection operator requires the existence of fluctuations for the resolved variables but also allows fluctuations for the unresolved variables, even though we did not utilize this in the current work. In power grid applications, if we treat the loads as unresolved variables, then we can envision scenarios where their initial conditions are allowed to fluctuate around some operational point. If we allow these fluctuations to be incorporated in the finite-rank projection operator used in the estimation of the memory, then we can obtain a representation of the memory (the usual bottleneck of reduced order models) which has built-in information about the fluctuations of the unresolved variables. This will be achieved while keeping the equations for the reduced order model to be valid {\it pathwise} and without the need for the introduction of a noise term as is usually required when fluctuations are allowed (see e.g. \cite{CHK00}). We note that while such a construction may incorporate basic aspects of the behavior of power grid systems, it is also interesting on its own and can be applied to other systems too.

Another interesting avenue to pursue is to investigate how the properties of the memory depend on the number of unresolved variables. For example, suppose we consider a power grid network with a few generators and a large number of loads. If we construct a reduced order model for the generators treating the loads as unresolved variables, an obvious question is whether the properties of the memory scale in some easily predictable fashion with the number of loads. This would allow to perform the expensive computations to obtain the memory for a small number of loads and then use the scaling relation to acquire the memory for a larger number of loads. Such a reduction in the computational cost of constructing the reduced model can bring the concept of model reduction closer to realistic applications. The existence of such a scaling would most likely need to assume a certain degree of homogeneity on the part of the loads. However, it is interesting to see if an approximate scaling exists even for inhomogeneous loads. 

These ideas are under investigation and we will report on them in a future publication.


\section*{Acknowledgments} We want to thank M. Anitescu and D. Barajas-Solano for useful discussions and comments. The work presented here was supported by the U.S. Department of Energy (DOE) Office of Science, Office of Advanced Scientific Computing Research (ASCR) as part of the Multifaceted Mathematics for Rare, Extreme Events in Complex Energy and Environment Systems (MACSER) project. Pacific Northwest National Laboratory is operated by Battelle for the DOE under Contract DE-AC05-76RL01830.


\bibliographystyle{plain}
\bibliography{refs}

\end{document}